\documentclass[12pt,a4paper]{article}
\usepackage{amsfonts,latexsym}
\usepackage{graphicx,color}
\usepackage{dcolumn}
\usepackage{graphicx}
\usepackage{amsmath}
\usepackage{amsfonts}
\usepackage{amssymb}
\usepackage{psfrag}
\usepackage{wrapfig}
\usepackage{subfigure}
\usepackage{makeidx}

\renewcommand{\thefootnote}{\fnsymbol{footnote}}

\begin{document}

\vspace{12mm}

\begin{center}
{{{\Large {\bf Scalarization of Einstein-Euler-Heisenberg black hole with multiple horizons }}}}\\[10mm]
{Hong Guo$^1$\footnote{e-mail address: hong\_guo@pku.edu.cn} and Yun Soo Myung$^{2}$\footnote{e-mail address: ysmyung@inje.ac.kr}}\\[8mm]
{${}^1$Kavli Institute for Astronomy and Astrophysics, Peking University, Beijing 100871, China\\[0pt]}
{${}^2${Center for Quantum Spacetime, Sogang University, Seoul 04107, Republic of  Korea\\[0pt] }}
\end{center}
\vspace{2mm}

\begin{abstract}
Scalarizations of the Einstein-Euler-Heisenberg (EEH) black hole  (EEHBH) with multiple horizons are investigated in the EEH-scalar theory by introducing a quadratic scalar coupling to the Maxwell term.
For mass $M=1$ and Euler-Heisenberg parameter $\mu=0.03$, the magnetically charged EEHBH admits four horizon families (low, cold, negative, and hot), with triple horizons appearing in the narrow band of magnetic charge $q\in[0.95,1.0065]$. 
The onset scalarization around the low, cold, and high horizons is then analyzed for the magnetic charge $q=0.5,\,1,\,2$, implying infinite branches of scalarized black holes for each case.
We construct the three fundamental branches of scalarized black holes. 
From the positivity condition of their mass, we find  the upper bounds on primary scalar charges $q_{s}$  for scalarized low and cold horizons. 
These bounds determine the allowable regions for the  Hawking temperature and entropy.
Furthermore, we perform a time-domain stability analysis and find that the instabilities arise only at small scalar charge regime.  Therefore,  stable and physically viable scalarized black holes exist in an intermediate window of the primary scalar charge for low and  cold horizon solutions and  a lower bound for hot horizon solution.

\end{abstract}

\vspace{1.5cm}

\hspace{11.5cm}
\newpage
\renewcommand{\thefootnote}{\arabic{footnote}}
\setcounter{footnote}{0}

\section{Introduction}
The black hole no-hair conjecture~\cite{Ruffini:1971bza}  has claimed that the final outcome of gravitational collapse with any types of matter is the Kerr-Newman black hole whose hairs  are solely given by mass, charge and angular momentum.  
Since the no-hair conjecture focuses on the fate of gravitational collapse but not the existence of black hole solutions with some types of matters, many black hole solutions with either new global charges or new non-trivial fields have been found over the past years. 
Bekenstein has formulated the no-scalar hair theorem~\cite{Bekenstein:1972ny,Bekenstein:1971hc} for black holes. This was based on three assumptions: i) A scalar is a canonically kinetic and minimally coupled field with a potential. ii) The scalar is a function of radial coordinate $r$ in the background of stationary black holes. iii) Its potential is everywhere non-negative.
Violation for one of these assumptions may lead to the evasion of no-scalar hair theorem~\cite{Herdeiro:2015waa}.

Later, an important mechanism was introduced to allow  the evasion of no-hair theorem~\cite{Bekenstein:1995un} through nonminimal scalar couplings to source terms that vanish asymptotically but induce a negative effective mass squared near the horizon.  
Two distinct classes of models have dominated the literature: curvature-induced scalarization, typically realized in Einstein-Gauss-Bonnet-Scalar (EGBS) theory ~\cite{Doneva:2017bvd,Silva:2017uqg,Antoniou:2017acq} and matter(charge)-induced scalarization explored in Einstein-Maxwell-Scalar (EMS) models~\cite{Herdeiro:2018wub, Myung:2018vug}.  
In the former, the Gauss-Bonnet invariant ($\mathcal{G}$) drives the tachyonic instability for Schwarzschild black hole, while in the latter the Maxwell term ($\mathcal{F}=F^2$) serves as a trigger to instability of Reissner-Nordstr\"om  black hole (RNBH). These led to spontaneous scalarization with infinite branches of scalarized black hole solutions. 

On the other hand, the Euler-Heisenberg (EH) Lagrangian has included a nonlinear electrodynamics term (NED: $\mathcal{F}^2$) to obtain a nonlinear extension of quantum electrodynamics (QED)~\cite{Heisenberg:1936nmg}.
In this case, the vacuum is regarded as a dynamically polarizable medium with polarization (magnetization) arising from virtual charge clouds around real charges (currents)~\cite{Obukhov:2002xa,Sorokin:2021tge}. 
Recently, Yajima and Tamaki have obtained the Einstein-Euler-Heisenberg black hole (EEHBH) solution described by ADM mass ($M$), charge ($q$), and the EH parameter ($\mu$) by considering the Einstein-Maxwell theory with a NED term (EEH theory)~\cite{Yajima:2000kw}. 
In the case of $\mu\le 0.08$ with $M=1$, it was shown that there exist multiple horizons depending on charge $q$, while for $\mu>0.08$, a single horizon appears as a function of charge $q$~\cite{Myung:2025zxu}. 

Concerning scalarization of EEHBHs, it is noted that the charged hairy black hole was obtained analytically  from the EEH-scalar (EEHS) theory together with a complicated scalar potential~\cite{Karakasis:2022xzm}.
More recently, a negative potential [$V(\phi) = -\alpha^2 \phi^6$]-induced scalarization of the EEHBH with single horizon ($\mu=0.3$) and unlimited charge $q$ was studied in the EEHS theory without nonminimal scalar coupling~\cite{Guo:2025ksj}. 
We observed a transition from primary to secondary scalar charge $q_s$ at $q=1/2$ with $M=1/2$. 
However, it turned out that the single branch of scalarized black holes is unstable against radial perturbations by computing quasinormal mode (QNM) frequency $\omega_I$ for the $s$-mode scalar.

On the other hand, spontaneous scalarization of the EEHBH with single horizon ($\mu=0.3$) was performed in the EEHS theory by introducing an exponential scalar coupling to the Maxwell term only~\cite{Zhang:2025msi}. 
In this case, infinite branches ($n=0,1,2,\cdots$) of  scalarized EEHBHs were found by taking into account the infinite scalar clouds.  
It is emphasized that the $n=0$ (fundamental) branch of scalarized EEHBHs is stable against radial perturbations, whereas the $n=1$ (excited) branch is unstable. 
Furthermore, considering two exponential scalar couplings to Maxwell and NED terms, spontaneous scalarization ($\alpha^+$ with $M=1$) of EEHBH was  available for charge $0<q< q_c$ with critical charge $q_c=1.115$ and positive coupling constant $\alpha$,  whereas its new scalarization ($\alpha^-$ with $M=1$) came out for $q> q_c$ and negative coupling constant $\alpha$~\cite{Zhang:2026bqu}.    
The former case with $q=0.5$ implies infinite branches of scalarized EEHBHs but its fundamental branch is stable against radial perturbations.  
On the other hand, the latter with $q=2,20$ shows two stable single branches of scalarized EEHBHs.   
These three cases might indicate features of  scalarization for EEHBHs with single horizon, compared to spontaneous scalarization for RNBH with double horizons in the EMS theory~\cite{Herdeiro:2018wub, Myung:2018vug}. 

In the present work,  we wish to perform scalarization of the EEHBH with multiple horizons in the EEHS theory by introducing a quadratic scalar coupling to the Maxwell term.
This black hole possesses triple horizons for a narrow range of $q\in[0.95,1.0065]$ when choosing black hole  mass $M=1$ and EH parameter $\mu=0.03<0.08$.  
The onset scalarization is carried out by choosing the low, (cold and negative), and high horizons for $q=0.5,1,2$, implying infinite branches of scalarized EEHBHs for low, cold, and high horizons because the negative horizon is very similar to the inner horizon of RNBH.   
Then, we construct three fundamental branches of (low, cold, high) scalarized EEHBHs and perform their stability analysis. 

\section{EEH black hole with multiple horizons}

The Einstein-Euler-Heisenberg-scalar (EEHS) action takes the form with the EH parameter $\mu$ to a nonlinear electrodynamics (NED) term $\mathcal{F}^2$
\begin{equation}
S_{\rm EEHS}=\frac{1}{16 \pi}\int d^4 x\sqrt{-g}\Big[ R-2\partial_\mu \phi \partial^\mu \phi-g(\phi)\mathcal{F}+\mu \mathcal{F}^2\Big],\label{Act1}
\end{equation}
where $g(\phi)=1-\alpha \phi^2$ is a scalar coupling function to the Maxwell term ($\mathcal{F}=F_{\mu\nu} F^{\mu\nu}$).
The Einstein equation is derived from the action (\ref{Act1})
\begin{eqnarray}
 G_{\mu\nu}=2\Big[\partial _\mu \phi\partial _\nu \phi -\frac{1}{2}(\partial \phi)^2g_{\mu\nu}+T_{\mu\nu}\Big] \label{equa1}
\end{eqnarray}
with its energy-momentum tensor
\begin{eqnarray}
T_{\mu\nu}&=&g(\phi)\Big(F_{\mu\rho}F_{\nu}~^\rho-\frac{1}{4}\mathcal{F}g_{\mu\nu}\Big)
          -2\mu \mathcal{F}\Big(F_{\mu\rho}F^\rho_\nu-\frac{1}{8}\mathcal{F}g_{\mu\nu}\Big) \label{emten}.
\end{eqnarray}
The Maxwell equation is given by
\begin{eqnarray} \label{M-eq}
&&\nabla_\mu (F^{\mu\nu}-2\mu\mathcal{F}F^{\mu\nu})=-g'(\phi)\nabla_{\mu} (\phi)F^{\mu\nu}.\label{M-eq1}
\end{eqnarray}
The scalar equation takes the simple form
\begin{equation}
\square \phi -\frac{\mathcal{F}}{4}g'(\phi)=0 \label{s-equa}.
\end{equation}
Introducing the mass function $\bar{m}(r)$ together with $\bar{A}_{\hat{\varphi}}=-q\cos\theta ~(\bar{\mathcal{F}}=\frac{2q^2}{r^4})$ and $\bar{\phi}=0$, the $({}_t~^t)$-component of the Einstein equation leads to a first-order equation for  $\bar{m}(r)$ as
\begin{equation}
\bar{m}'(r)=\frac{q^2}{2r^2}-\mu \frac{q^4}{r^6}. \label{mass-eq}
\end{equation}
Solving Eq.(\ref{mass-eq}) leads to the EEH black hole (EEHBH) solution~\cite{Yajima:2000kw,Amaro:2020xro,Breton:2021mju,Allahyari:2019jqz}
\begin{eqnarray}
ds^2_{\rm EEH}=\bar{g}_{\mu\nu}dx^\mu dx^\nu=-f(r) dt^2+\frac{dr^2}{f(r)} +r^2d\Omega^2_2   \label{EEH-s}
\end{eqnarray}
with the metric function
\begin{equation} \label{metric-func}
 f(r)\equiv1-\frac{2\bar{m}(r,M,q,\mu)}{r}=1-\frac{2M}{r}+\frac{q^2}{r^2}-\frac{2\mu q^4}{5r^6}.
\end{equation}
This black hole solution is described by three parameters $(M,q,\mu)$ where $M$ denotes the ADM mass, $q$ is the magnetic charge, and $\mu$ is the EH parameter to control the NED term.
In the limit of $\mu\to 0$, one recovers the RNBH metric function
\begin{equation}
f_{RN}(r)=1-\frac{2M}{r}+\frac{q^2}{r^2}
\end{equation}
whose two outer/inner horizons are given by
\begin{equation}
r_{RN\pm}=M^2\pm \sqrt{M^2-q^2}, \quad r_{RN+}(M,q)\to r_{RN}(M,q).
\end{equation}

\begin{figure*}[t!]
   \centering
    \includegraphics[width=0.4\textwidth]{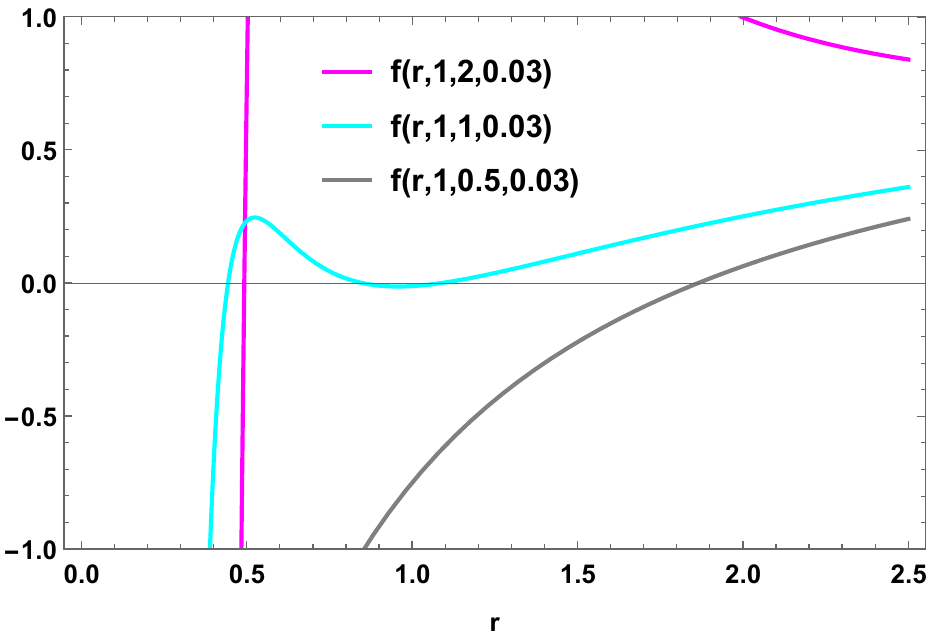}
   \hfill%
   \includegraphics[width=0.4\textwidth]{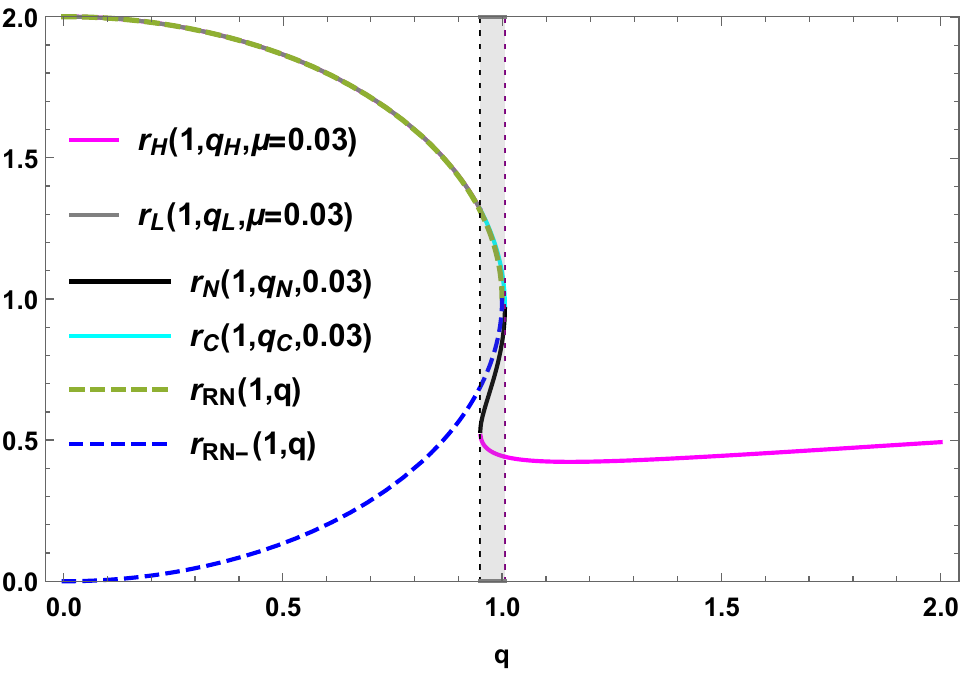}
\caption{(Left) Solution to $f(r,M=1,q,0.03)=0$ with $q=0.5,1,2$. For $q=1$, one finds three real roots, implying three horizons.
 (Right) Four horizons  $r_{L}(M=1,q_L,\mu),~r_{C}(1,q_C,\mu),~ r_{N}(1,q_N,\mu),~ r_{H}(1,q_H,\mu)$  as functions of $q$ with $\mu=0.03$. $r_{RN}(M=1,q\in[0,1])$ represents the outer horizon for RN black hole, while $r_{RN-}(1,q)$ denotes its inner horizon. The shaded region denotes a narrow region of $q\in[0.95,1.0065]$, where triple horizons exist.}\label{fig1}
\end{figure*}

Hereafter, we wish to choose the case of $\mu\le 0.08$ to get a black hole  with multiple horizons.  
In the case of $\mu= 0.03$ with $M=1$~\cite{Myung:2025zxu}, there exist four horizons  which make scalarization analysis complicated. 
As is shown in (Left) Fig.~\ref{fig1}, one finds that for $q=1$, three real roots exist, implying three horizons.
From $f(r)=0$ with $\mu=0.03$, we need four horizons (positive roots) to cover $q>0$ [see (Right) Fig.~\ref{fig1}] as
\begin{equation}
r_{L}(M,q_L,\mu),\quad r_{H}(M,q_H,\mu),\quad r_{C}(M,q_C,\mu),\quad r_{N}(M,q_N,\mu),
\end{equation}
whose forms are too complicated to show here. 
Here, $q_i$ is labeled as: $q_L\in[0,0.95],~q_H\in[0.95,\infty],~q_C\in[0.95,1.0065],$ and $q_N\in[1.0065,0.95]$. 
Triple horizons of $r_H,~r_C,~r_N$ exist only for a narrow region of $q\in[0.95,1.0065]$. 
We note that $r_L$ and $r_H$ are connected to each other for $\mu>0.08$, but they are disconnected for $\mu\le0.08$.  
One dotted line represents the line at $q=0.95$ where the low horizon meets the cold one and the negative horizon meets the hot one, while the other denotes $q=1.0065$ (extremal BH) where the cold horizon meets the negative horizon.  
The names of the subscripts become clear when their thermodynamic quantities are found: L, H, N, C denote low, hot, negative, cold, respectively.  
We emphasize that $r_{N}$ resembles the inner horizon of RNBH and thus, it may cause some trouble for scalarization analysis. 

In addition, this bald black hole might be found from the infinite series of Maxwell terms $L_{EM}=\sum_{k=1}^\infty \alpha_k \mathcal{F}^k$~\cite{Gao:2021kvr}.  For $\alpha_1=1,~\alpha_2(=\mu),~\alpha_3=4\alpha_2^2,~\alpha_4=24\alpha_2^3,~\alpha_5=176\alpha_2^4,\cdots$, one found the EEHBH with triple horizons for $M=1,~Q=0.7,~\alpha_2=-0.001$.

We would like to note that this bald black hole is similar to nonlinear scalarized black holes with two horizons found from the EMS theory with a quartic coupling function $\tilde{g}(\phi)=1+\alpha \phi^4$ and an electric charge $Q$: the hot and the cold branches with the existing bald (RNBH) branch~\cite{Blazquez-Salcedo:2020nhs,LuisBlazquez-Salcedo:2020rqp}. 
It is worth noting that the hot branch is stable against radial perturbations, while the cold branch is unstable. 

\begin{figure*}[t!]
   \centering
  \includegraphics[width=0.4\textwidth]{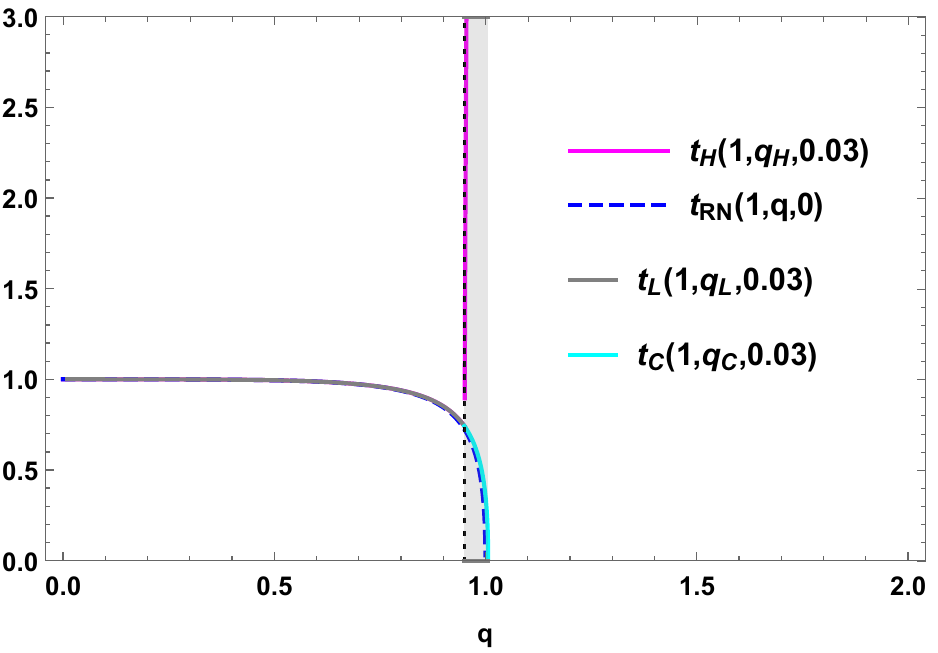}
 \hfill%
    \includegraphics[width=0.4\textwidth]{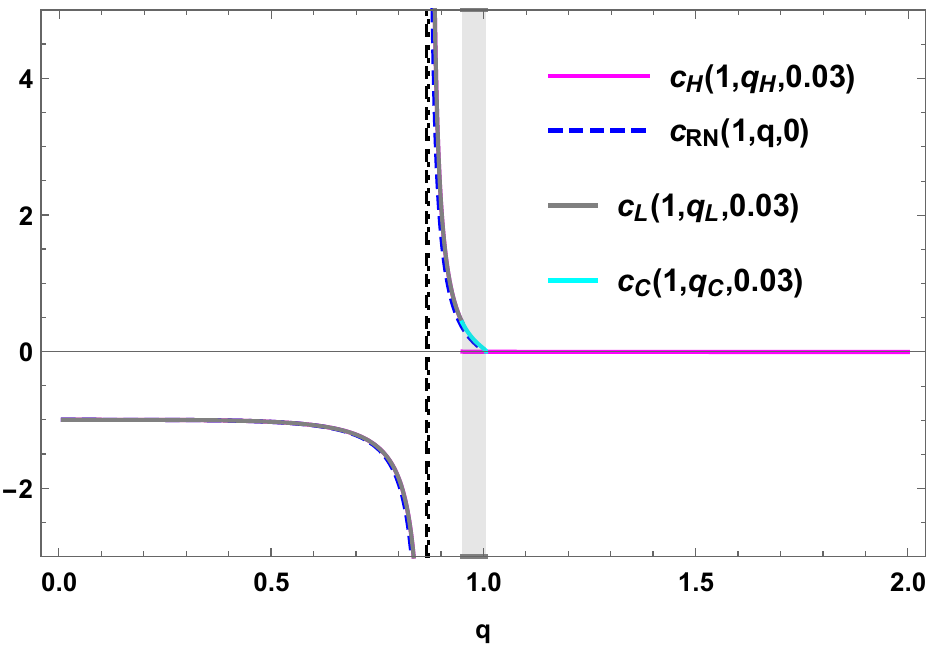}
\caption{(Left) Three reduced temperatures $t_i(M=1,q_i,\mu=0.03)$ for $i=L,C,H$ with $t_{RN}(1,q)$. A dotted line denotes a line at $q=0.95$ which three horizons meet.
(Right) Three reduced heat capacities  $c_i(M=1,q_i,\mu=0.03)$ for $i=L,C,H$ with $c_{RN}(1,q)$. 
It includes two Davies points [dashed lines at $q=0.866(RN),0871(L)$] and two extremal points with $t_{C}(1,1.0065)=c_{C}(1,1.0065)=0$ and $t_{RN}(1,1)=c_{RN}(1,1)=0$. The shaded region represents a narrow region of $q\in [0.95,1.0065]$.}\label{fig2}
\end{figure*}

Also, the reduced Hawking temperature (surface gravity) and reduced heat capacity are given by~\cite{Myung:2025zxu}
\begin{eqnarray}
t_{i}(M,q_i,\mu)&=&4\kappa=2f'(r_i)=\frac{2}{r_{i}(M,q_i,\mu)}\Big[1- \frac{q_i^2}{r_{i}^2(M,q_i,\mu)}-\frac{2\mu q_i^4}{r_{i}^6(M,q_i,\mu)}\Big], \label{sg1}\\
c_i(M,q_i,\mu)&=&-\frac{r_i^2(M,q_i,\mu)}{4}\Big[\frac{2\mu q_i^4-q_i^2 r_i^4(M,q_i,\mu)+r_i^6(M,q_i,\mu)}{14\mu q_i^4-3q_i^2r_i^4(M,q_i,\mu)+r_i^6(M,q_i,\mu)}\Big] \label{sg2}
\end{eqnarray}
with $i=L,H,C$.
On the other hand, the corresponding quantities for RNBHs are given by
\begin{eqnarray}
t_{RN}(M,q)&=&\frac{2}{ r_{ RN}(M,q)}\Big[1-\frac{q^2}{r_{RN}^2(M,q)}\Big], \label{RN-t} \\
c_{RN}(M,q)&=&-\frac{r_{RN}^2(M,q)}{4}\Big[\frac{r_{RN}^2(M,q)-q^2}{r_{ RN}^2(M,q)-3q^2}\Big].\label{RN-hc}
\end{eqnarray}
(Left) Fig.~\ref{fig2} shows clearly why we use the subscripts of L, H, C to represent low, high, cold temperatures, respectively.
(Right) Fig.~\ref{fig2} indicates heat capacity that L-horizon includes Davies point where a rapid phase transition occurs. 
We note that C-horizon is thermodynamically stable, whereas the L- and H-horizons include  thermodynamically unstable regions.   
Thermodynamically stable regions are specified as:    $q\in[0.871,0.95]$ for the L-horizon, $q\in[0.95,1.0065]$ for the C-horizon, and $q\in[0.866,1]$ for the RNBH-horizon.  
However, we do not want to display the N-horizon with negative temperature because its property is very similar to the inner horizon $r_{RN-}(1,q)$ for RNBH.

\section{Linearized analysis}

Now, we are in a position to introduce perturbations around the EEHBH background as
\begin{equation}
g_{\mu\nu}=\bar{g}_{\mu\nu}+h_{\mu\nu},\quad\phi=0+\delta \varphi, \quad F_{\mu\nu}=\bar{F}_{\mu\nu}+f_{\mu\nu},\quad f_{\mu\nu} =\partial_\mu a_\nu-\partial_\nu a_\mu.
\end{equation}
It is worthy to note that the linearized EEH theory around electrically charged EEHBH leads to being stable against the metric-vector perturbations~\cite{Luo:2022gdz}.
Hence, we focus on solving the linearized scalar equation which determines the tachyonic instability of EEHBH
\begin{equation}
\Big[\bar{\square}-m^2_{\rm eff}\Big]\delta \varphi=0,\quad m^2_{\rm eff}(r,q)=-\alpha \frac{q^2}{r^4}.\label{per-eq}
\end{equation}
Let us introduce a tortoise coordinate defined by $dr_*=\frac{dr}{f(r)}$ and consider a separation of variables
\begin{equation}
\delta\phi(t,r_*,\theta,\varphi)=\sum_m\sum^\infty_{l=|m|}\frac{\varphi_{lm}(t,r_*)}{r}Y_{lm}(\theta,\varphi).
\end{equation}
Its $s(l=0,m=0)$-mode linearized equation reduces to
\begin{equation} \label{mode-d}
\frac{\partial^2\varphi_{00}(t,r_*)}{\partial r_*^2} -\frac{\partial^2\varphi_{00}(t,r_*)}{\partial t^2}=V_{\rm EEH}(r)\varphi_{00}(t,r_*),
\end{equation}
where the $s(l=0)$-mode potential is given by
\begin{eqnarray}
V_{\rm EEH}(r,M,q,\alpha)&=&f(r)\Big[\frac{2M}{r^3}-\frac{2q^2}{r^4}+\frac{12\mu q^4}{5r^8}+m^2_{\rm eff}\Big].\label{EEH-P}
\end{eqnarray}
Now we may introduce the scalar potentials around $i$-horizons
\begin{eqnarray}
&&V_{i{\rm EEH}}(r,M,q_i,\alpha^i)=f_i(r)\Big[\frac{2M}{r^3}-\frac{2q_i^2}{r^4}+\frac{12\mu q_i^4}{5r^8}+m^2_{i,{\rm eff}}\Big]\label{IEEH-P}
\end{eqnarray}
with
\begin{equation} \label{If-fun}
f_i(r)=1-\frac{2M}{r}+\frac{q_i^2}{r^2}-\frac{2\mu q_i^4}{5r^6},\quad m^2_{i,{\rm eff}}=-\frac{\alpha^i q^2_i}{r^4}.
\end{equation}

\begin{figure*}[t!]
\centering
\includegraphics[width=0.4\textwidth]{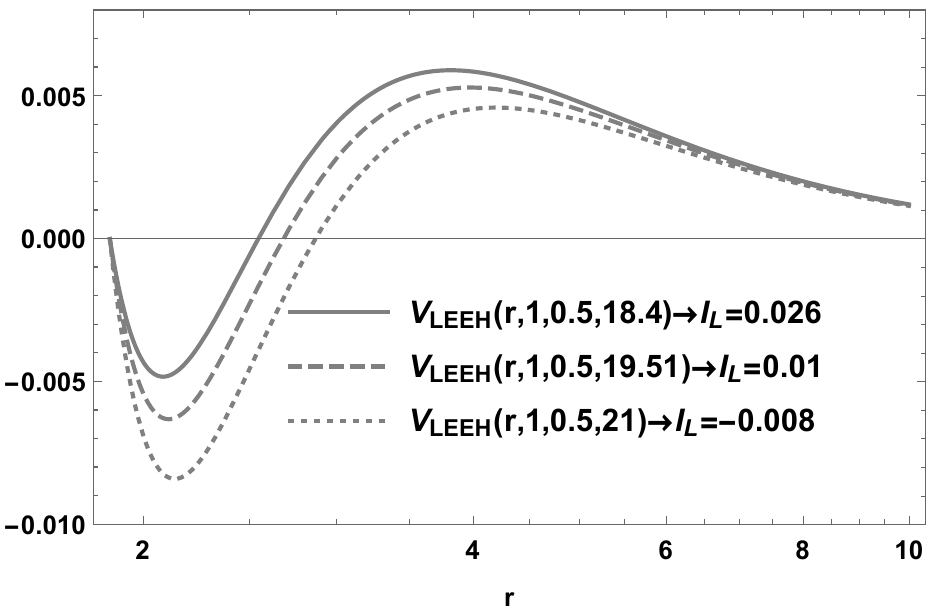}
\hfill%
\includegraphics[width=0.4\textwidth]{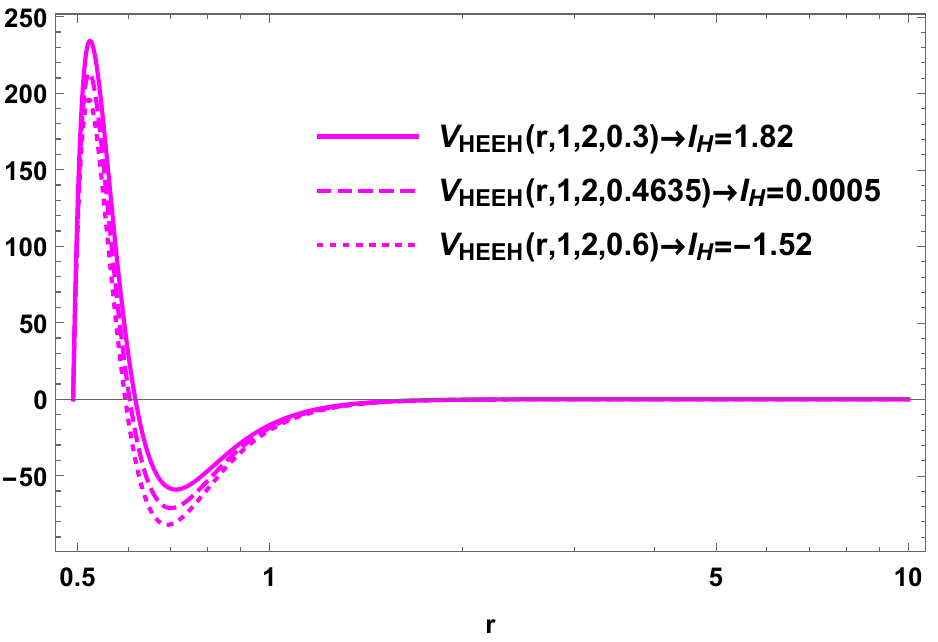}
\caption{Scalar potential  $V_{i{\rm EEH}}(r,M=1,q,\alpha)$ and its integration $I_i$ for  $l = 0$-scalar mode. (Left) $q_L=0.5$ for $r\in[r_L=1.87,10]$. Here, $\alpha^L$ takes 18.4, 19.51, 21. (Right)  $q_H=2$ for $r\in[r_H=0.493,10]$.  Here, $\alpha^H$ is chosen as  0.3, 0.4635, 0.6.}\label{fig3}
\end{figure*}

\begin{figure*}[t!]
\centering
\includegraphics[width=0.3\textwidth]{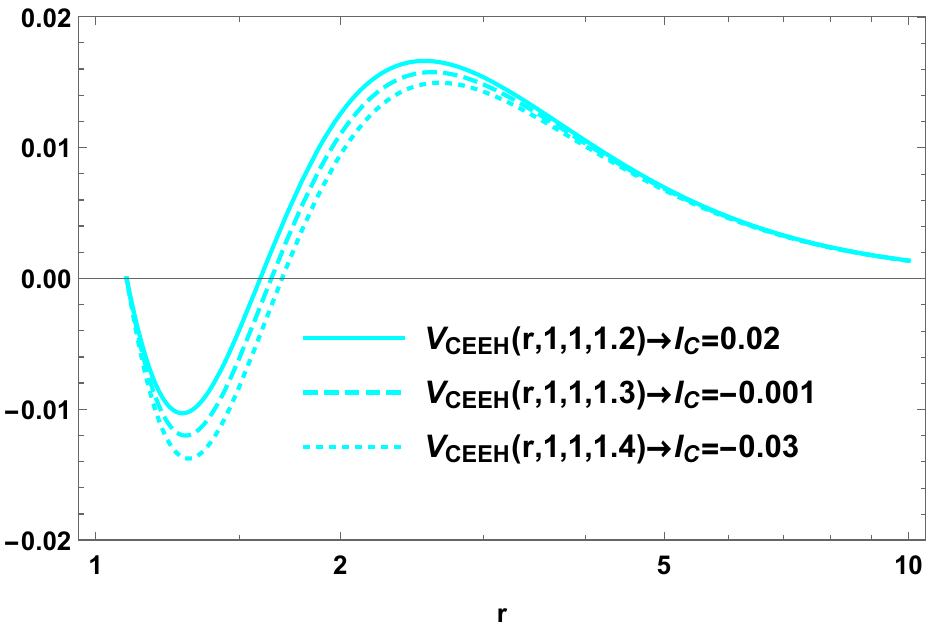}
\hfill%
\includegraphics[width=0.3\textwidth]{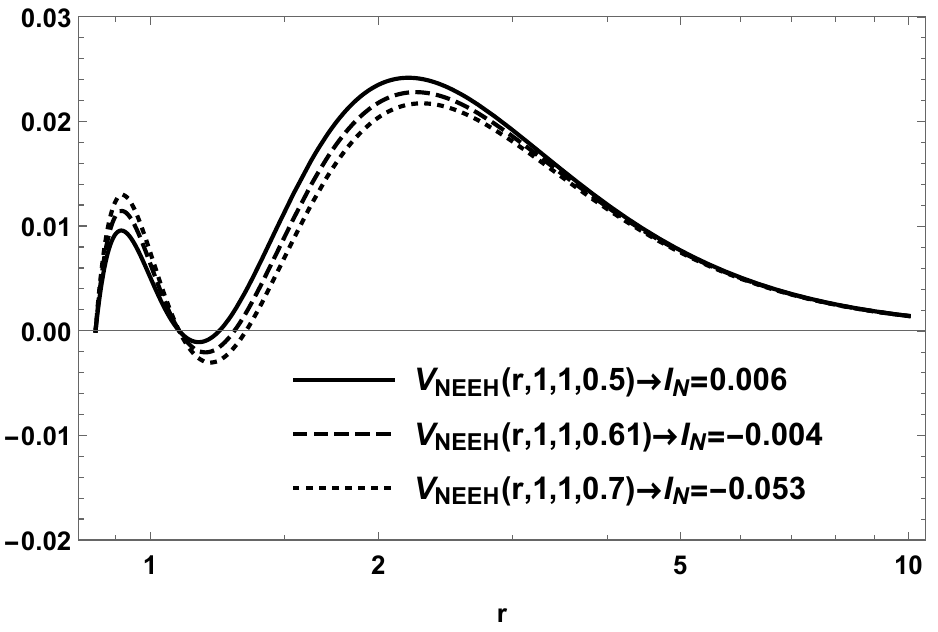}
\hfill%
\includegraphics[width=0.3\textwidth]{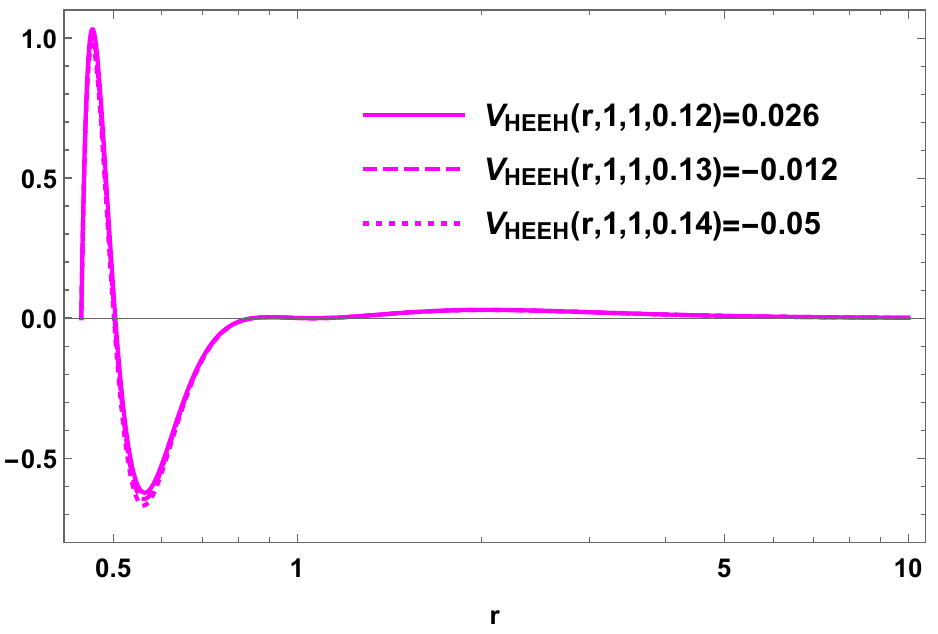}
\caption{ Three Scalar potentials  $V_{i{\rm EEH}}(r,M=1,q=1,\alpha^i)$ and their integration $I_i$ for  $l = 0$-scalar mode in the narrow band of $q(=1)\in[0.95,1.0065]$. (Left) $r\in[r_C=1.092,10]$.  (Middle)  $r\in[r_N=0.8475,10]$. (Right)  $r\in[r_H=0.4438,10]$. }\label{fig4}
\end{figure*}

First of all, it is easy to compute the sufficient condition for tachyonic instability given by~\cite{Dotti:2004sh}
\begin{equation}
\int_{r_i(M,q_i,\mu)}^\infty \Big[\frac{V_{i{\rm  EEH}}(r)}{f_i(r)}\Big]dr\equiv I_i<0
\end{equation}
which determines $\alpha_{s,i}(M,q_i,\mu)$ as the sufficient instability condition for $i$-horizons with $g_i(\phi)=1-\alpha^i \phi^2$.

\begin{figure*}[t!]
   \centering
  \includegraphics[width=0.4\textwidth]{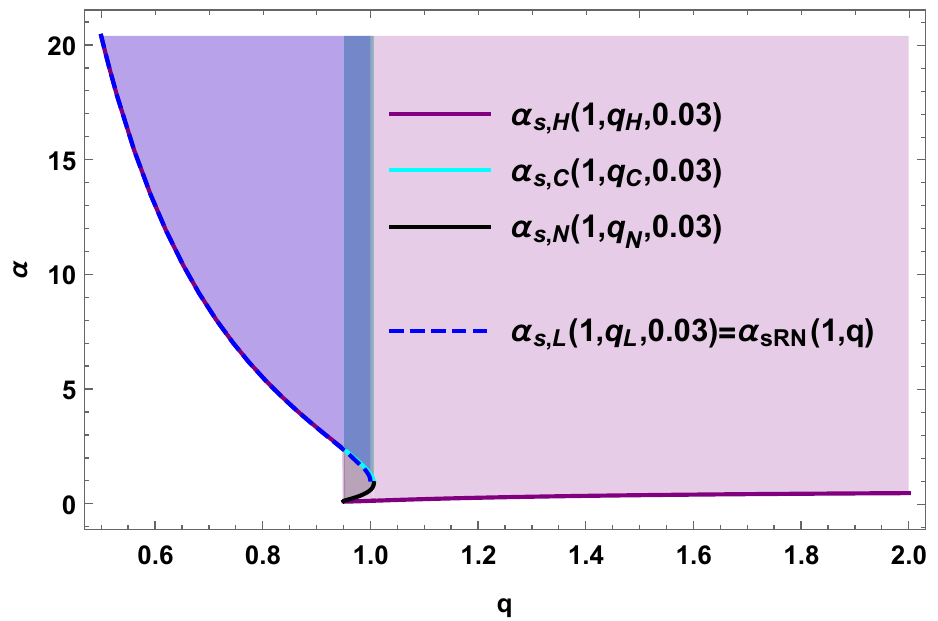}
   \hfill%
\includegraphics[width=0.4\textwidth]{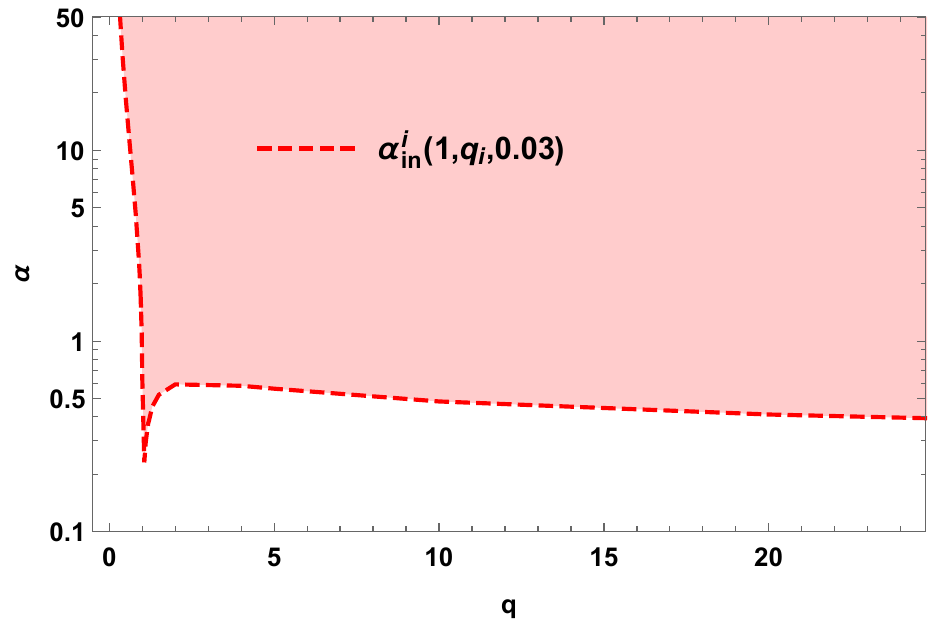}
\caption{(Left) Sufficient conditions for instability $\alpha_{ s,i}(1,q_i,0.03)$ and $\alpha_{\rm sRN}(1,q)$, leading to $\alpha_{s,L}(1,q_L,0.03)\approx\alpha_{\rm sRN}(1,q)$ for $q\in[0,0.95]$.
 The whole shaded region represents unstable region of $\alpha(1,q_i,0.03)\ge \alpha_{s,i}(1,q_i,0.03)$. (Right) Instability condition $\alpha(1,q_i,0.03)\ge \alpha_{in}^{i}(1,q_i,0.03)$ for $i=L,~C,~H$.  }\label{fig5}
\end{figure*}

We present explicit forms $\alpha_{ s,i}(M,q_i,\mu=0.03)$ and $\alpha_{\rm sRN}(M,q)$~\cite{Myung:2018vug} obtained from $I_i=0$ and $I_{RN}=0$ as
\begin{eqnarray}
\alpha_{s,i}(M,q_i,0.03)&=&-2+\frac{0.0308571~ q_i^2}{ r_i^4(M,q_i,0.03)}+\frac{3r_i(M,q_i,0.03)}{q_i^2}, \label{alp-1} \\
 \alpha_{sRN}(M,q)&=&-2+\frac{3r_{RN}(M,q)}{q^2}. \label{alp-2}
\end{eqnarray}
Fig.~\ref{fig3} shows potential $V_{i{\rm EEH}}(r,M=1,q_i,\alpha^i)$ and its integration $I_i$ for $l = 0$-scalar mode with single horizons $r_L$ and $r_H$.
For $q=0.5$, its sufficient instability occurs at  $\alpha_{s,L}(M=1,q=0.5,0.03)=20.39$ and for $q=2$, it is given by $\alpha_{s,H}(1,2,0.03)=0.4635$.
We observe that the $q=2$ potential has an upside-down form of the $q=0.5$ potential.
For $q=1$ (Fig.~\ref{fig4}), there are three different potentials for triple horizons in the narrow band of $q(=1)\in[0.95,1.0065]$. 
Accordingly, we obtain three different sufficient conditions for instability: $\alpha_{s,C}(1,1,0.03)=1.3,~\alpha_{s,N}(1,1,0.03)=0.61,~\alpha_{s,H}(1,1,0.03)=0.13$.
One finds from (Left) Fig. 5 that $\alpha_{s,L}(1,q_L\in[0,0.95],0.03)\approx \alpha_{ sRN}(1,q\in[0,1])$.

To determine the instability condition $\alpha^i_{in,n}(M,q_i,\mu)$, the WKB integral is approximated by considering the last term in \eqref{EEH-P} as~\cite{Hod:2019ulh}
\begin{equation}
\int^{\infty}_{r_i}V_{b,i}(r)dr= \sqrt{\alpha^i}\cdot  q_i\int^{\infty}_{r_i(M,q_i,\mu)} \frac{dr}{r^2\sqrt{f_i(r)}}\equiv\sqrt{\alpha^i} I^i_{n}(M,q_i,\mu) =\Big(n_i+\frac{3}{4}\Big)\pi,
\end{equation}
which could be integrated numerically to yield
\begin{equation} \label{alphan}
\alpha^i_{ in,n}(M,q_i,\mu)=\Big[\frac{\pi(n_i+3/4)}{I^i_n(M,q_i,\mu)}\Big]^2,\quad  n_i=0,1,2,\cdots.
\end{equation}

\begin{figure*}[t!]
   \centering
  \includegraphics[width=0.4\textwidth]{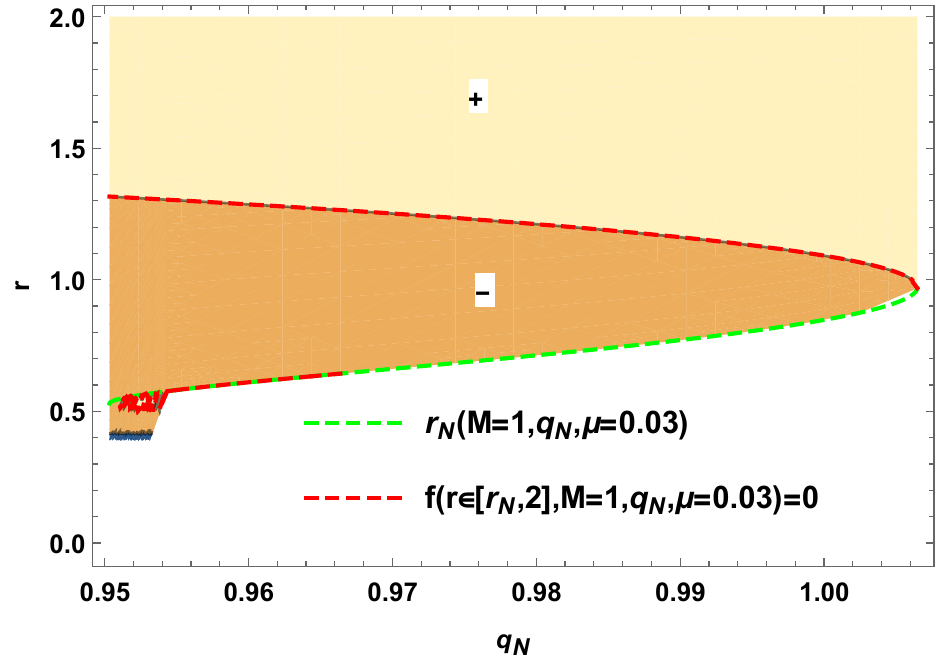}
   \hfill%
\includegraphics[width=0.4\textwidth]{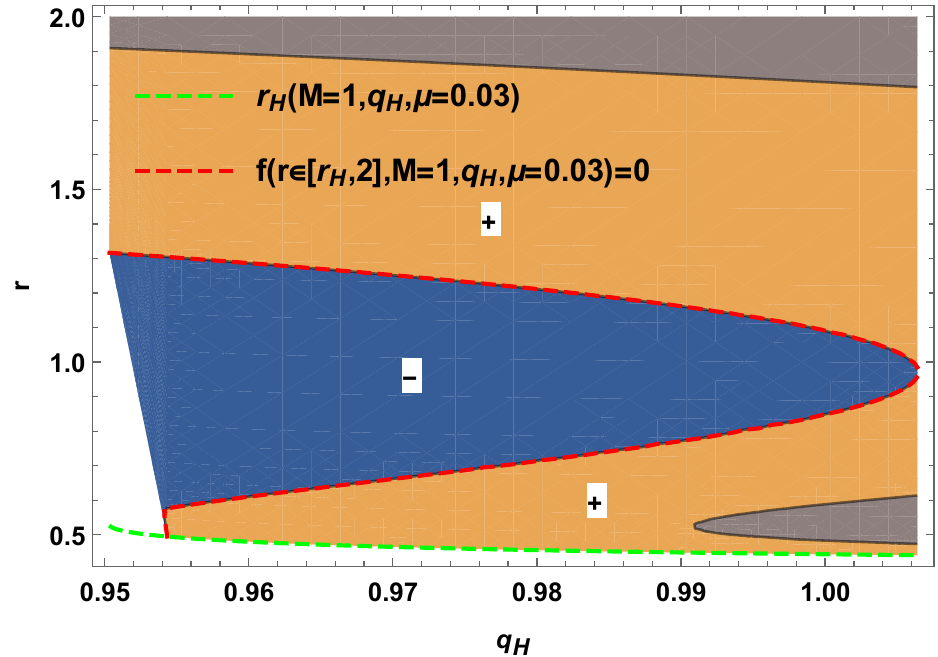}
\caption{(Left) Negative region of the metric function  $f_H(r\in[r_H,2],M=1,q_H\in[0.95,1.0065],\mu=0.03)$ in the near-horizon. (Right)  Negative region of the metric function  $f_N(r\in[r_N,2],M=1,q_N\in[0.95,1.0065],\mu=0.03)$ in the near-horizon.   }\label{fig6}
\end{figure*}

We plot $\alpha^i_{ in}(M=1,q_i,0.03)[\equiv\alpha^i_{\rm in,n=0}(1,q_i,0.03)]$ from numerical data in (Right) Fig.~\ref{fig5}. 
Here, we have $\alpha^C_{in}(1,1,0.03)=0.62$, but $\alpha^N_{in}(1,1,0.03)$ is not properly defined. 
Also $\alpha^H_{in}(1,1,0.03)$ is not defined, but one has $\alpha^H_{in}(1,q=1.06,0.03)=0.23$ and $\alpha^H_{in}(1,q=1.1,0.03)=0.29$.
Others of $\alpha^i_{in,n\not=0}(1,q_i,0.03) $ may be  used to estimate higher branch points for i-horizons.
For example, it predicts three cases: for $q_L=0.5$,  $\alpha^L_{ in,0}=18.4<\alpha_{s,L}$,  $\alpha^L_{ in,1}=100 $, $\alpha^L_{in,2}=247$; for $q_C=1$, $\alpha^C_{in,0}=0.62>\alpha_{s,C}$,  $\alpha^C_{ in,1}=3.38$,  $\alpha^C_{ in,2}=8.34$; for $q_H=2$,  $\alpha^H_{in,0}=0.59>\alpha_{s,H} $, $\alpha^H_{in,1}=3.20 $, $\alpha^H_{in,2}=7.91$.
Here, we wish to point out that the H, N-integrations are not properly defined because their metric function  $f_i(r)$ is negative in the near-horizon (see Fig.~\ref{fig6}).
This suggests that the C-horizon is regarded as a representative for the narrow region of $q\in[0.95,1.0065]$ to carry scalarization out.

Finally, we may solve the static linearized equation  to obtain the threshold of instability $\alpha^i_{th}(1,q_i,0.03)$ directly.
Also, this computation can provide the scalar seeds $\{\varphi^i_n(r)\}$ for $n=0,~ 1,~ 2,\cdots$ branches of scalarized i-horizons.
For this purpose, we introduce the static linearized equation for $s$-mode $\varphi^i(r)$ around the i-horizon
\begin{equation} \label{ssclar-eq}
\frac{1}{r^2}\Big[r^2f_i(r)(\varphi^i)'(r)\Big]'+\frac{\alpha^i q_i^2}{r^4} \varphi^i(r)=0,
\end{equation}
which describes an eigenvalue problem at the i-horizon. 
Requiring an asymptotically vanishing scalar [$\varphi^i(r\to \infty)=0$] implies that a smooth scalar selects a discrete set of node numbers $n=0$, 1, 2, $\cdots$ for the i-horizon.   
Also, it determines the bifurcation points [$\alpha^i_n(1,q_i,0.03)$] precisely. 
In this case, it may be confirmed that $\alpha^i_{ th}(1,q_i,0.03)=\alpha^i_{n=0}(1,q_i,0.03)$ for $i=L,~C,~H$.
We note that the scalar seed $\varphi^L_0(r)$ without zero crossing will develop the scalar hair $\phi_L(r)$ which describes the $n=0$ fundamental branch of scalarized L-horizon for $\alpha^L\ge \alpha^L_0(=\alpha^L_{ th})$.
To each i-horizon, infinite branches ($n=0,~1,~2,\cdots$) can be constructed from infinite $\alpha^i$-bounds: $\alpha^i\ge \alpha^i_0,~\alpha^i\ge \alpha^i_1$, $\alpha^i\ge \alpha^i_2,~\cdots$. 
It may describe briefly a prescription for how to obtain infinite branches of scalarized i-horizons  through spontaneous scalarization.

\section{Scalarized i-horizon solutions }

One expects that infinite branches of scalarized i-horizons could be generated from the scalar clouds $\{\varphi^i_n(r)\}$ for $i=L,H,C$ in the unstable region of EEHBHs [$\alpha^i(1,q_i,0.03)\ge \alpha^i_{th}(1,q_i,0.03)$].
However, it is unlikely that infinite branches of scalarized N-horizon can be found from the scalar cloud  $\{\varphi^N_n(r)\}$ because the latter is ill-defined.

We wish to derive scalarized i-horizons by solving full equations.
For this purpose, we introduce the metric and fields as~\cite{Herdeiro:2018wub}
\begin{eqnarray}\label{nansatz}
ds^2_{i}&=&-N_i(r)e^{-2\delta_i(r)}dt^2+\frac{dr^2}{N_i(r)}+r^2(d\theta^2+\sin^2\theta d\hat{\varphi}^2) \nonumber \\
N_i(r)&=&1-\frac{2m_i(r)}{r},\quad \phi_i=\phi_i(r),\quad A= A_{\hat{\varphi}} d\hat{\varphi}.
\end{eqnarray}
Substituting the gauge field ansatz into Eq.(\ref{M-eq}), one finds a magnetic potential $A_{\hat{\varphi}}=-q \cos \theta$, leading to $F_{\theta \hat{\varphi}}=q\sin\theta$ and $\mathcal{F}=2q^2/r^4$.
This means that it completes solving the Maxwell part. 
Plugging \eqref{nansatz} into Eqs.~\eqref{equa1} and \eqref{s-equa}, three equations for $m_i(r),~\delta_i(r),~\phi_i(r)$ are given by
\begin{align}
&\Big(q_i^2-\frac{2\mu q_i^4}{r^4}\Big)-\alpha^i q_i^2\phi_i^2-2r^2 m_i'(r)+r^3\Big(r-2m_i(r)\Big)\phi_i'^2(r)=0, \label{neom1}\\
&\delta_i'(r)+r\phi_i'^2(r)=0, \label{neom2}\\
&\alpha^i q_i^2\phi_i(r)-2r^2\Big[m_i(r)+rm_i'(r)-r\Big]\phi_i'(r) -r^3\Big(r-2m_i(r)\Big)\Big[\delta_i'(r)\phi_i'(r)-\phi_i''(r)\Big]=0. \label{neom3}
\end{align}
Here, the prime ($'$) denotes differentiation with respect to $r$. 
It is useful to note that Eq.(\ref{neom1}) reduces to Eq.(\ref{mass-eq}) for $\phi_i(r)=0$.

Guaranteeing the existence of a horizon located at $r=r_i$, an approximate solution to Eqs.\eqref{neom1}-\eqref{neom3} takes the form in the near-horizon as
\begin{eqnarray}
m_i(r)&=&\frac{r_i}{2}+m_{1,i}(r-r_i)+\cdots,\label{aps-0} \\
\delta_i(r)&=&\delta_{0,i}+\delta_{1,i}(r-r_i)+\cdots,\label{aps-1}\\
\phi_i(r)&=&\phi_{0,i}+\phi_{1,i}(r-r_i)+\cdots,\label{aps-2}
\end{eqnarray}
where the three coefficients are given by
\begin{eqnarray}\label{ncoef1}
&&m_{1,i}=\frac{q_i^2(1-\alpha^i\phi^2_{0,i})}{2r_i^2}-\frac{\mu q_i^4}{r_i^6},\quad \delta_{1,i}=-r_i\phi_{1,i}^2,\quad \phi_{1,i}=\frac{\alpha^i q_i^2 \phi_{0,i}}{r_i^3(m_{1,i}-1)}.
\end{eqnarray}
Here, two horizon parameters of $\phi_{0,i}=\phi(r_i,\alpha^i)$ and $\delta_{0,i}=\delta_i(r_i,\alpha^i)$ will be determined when matching with an asymptotically flat solution in the far-region
\begin{eqnarray}\label{ncoef2}
&&m_i(r)=M_i-\frac{q_i^2+q_{s,i}^2}{2r}+\cdots,\quad
\delta_i(r)=\frac{q_{s,i}^2}{2r^2}+\cdots,\quad
\phi_i(r)=\frac{q_{s,i}}{r}+\cdots,
\end{eqnarray}
where $q_{s,i}$ represents the primary scalar charge for the i-horizon, in addition to the ADM mass $M_i$, and the magnetic charge $q_i$.
It is emphasized that the function $\delta(r)$ itself is not gauge invariant, since a constant rescaling of the time coordinate ($t \rightarrow \tilde t = e^{\delta_0} t$) induces the shift ($\delta(r)\rightarrow \delta(r)-\delta_0$).
Hence, we adopt a gauge choice of $\delta_i(\infty)=0$, which corresponds to the standard normalization of the time coordinate at spatial infinity. 
A different choice of $\delta_i(r_h)=0$ at the event horizon is equally valid and is physically equivalent to the present convention. 
Two descriptions are related by a redefinition of the time coordinate and all physical observables remain unchanged under this transformation.

\begin{figure*}[t!]
\centering
\includegraphics[width=0.3\textwidth]{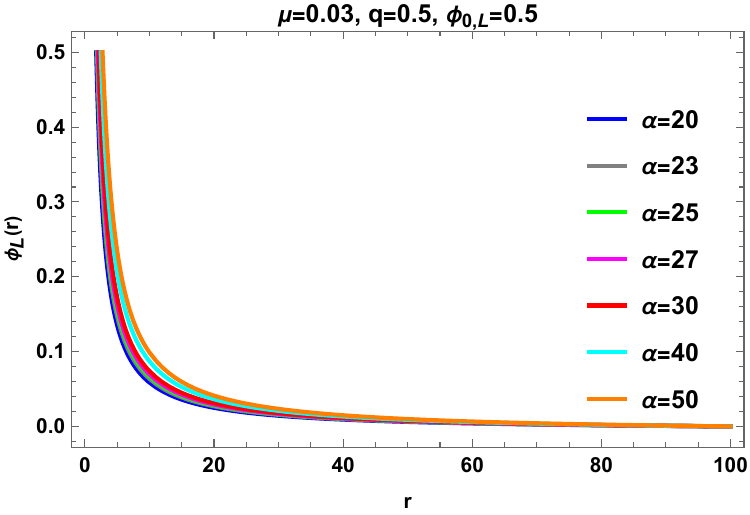}
\hfill%
\includegraphics[width=0.3\textwidth]{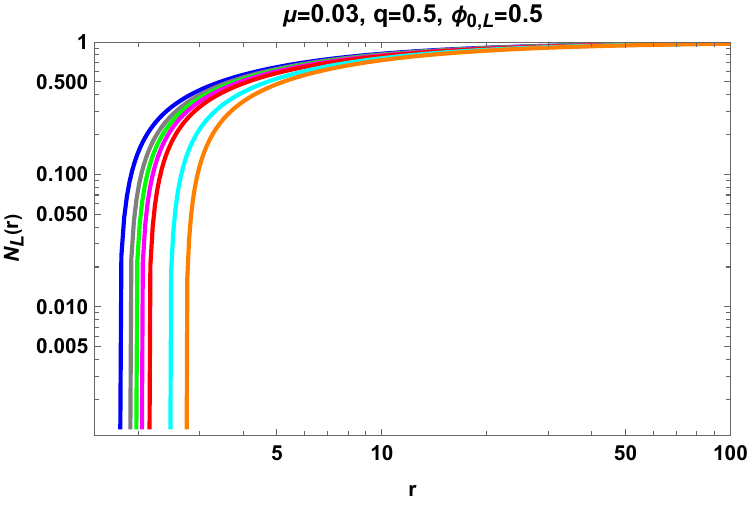}
\hfill%
\includegraphics[width=0.3\textwidth]{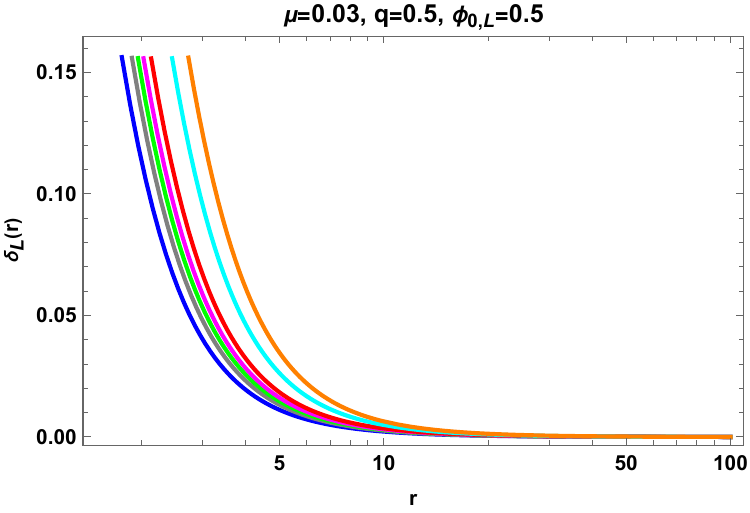}
\hfill%
\includegraphics[width=0.3\textwidth]{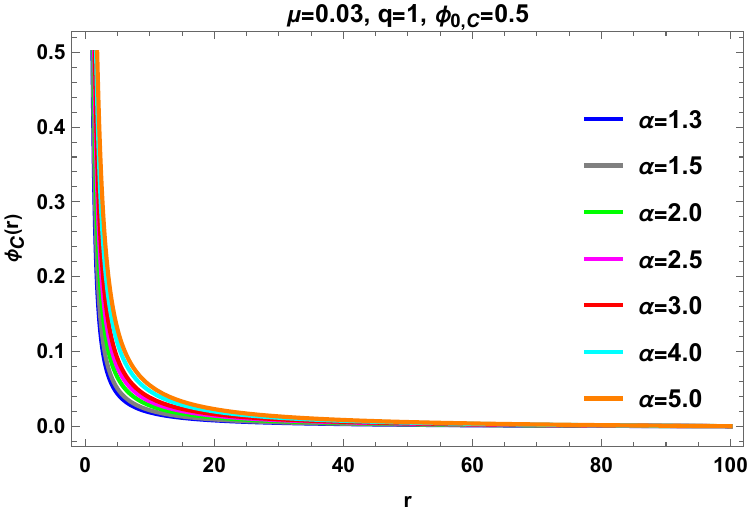}
\hfill%
\includegraphics[width=0.3\textwidth]{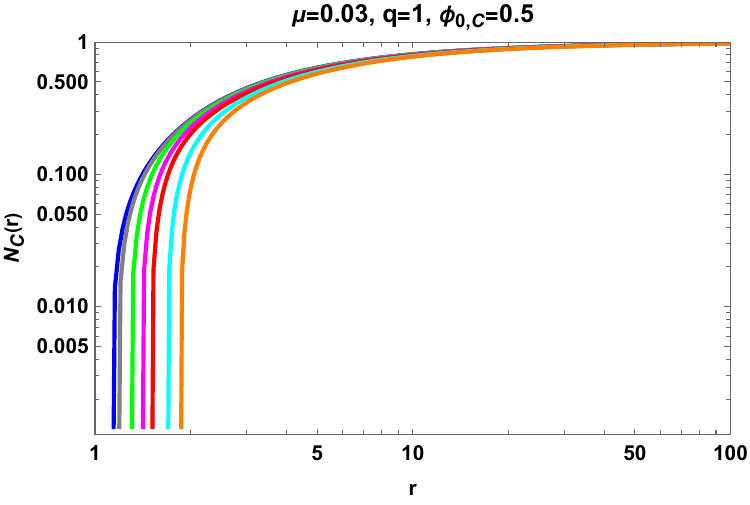}
\hfill%
\includegraphics[width=0.3\textwidth]{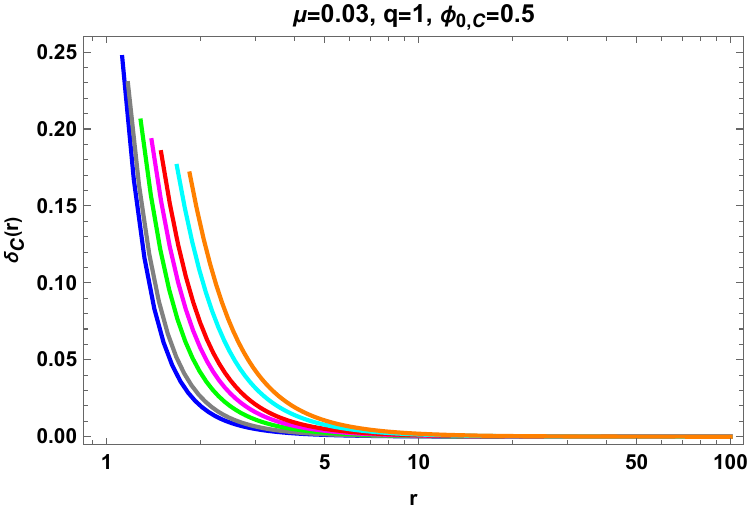}
\hfill%
\includegraphics[width=0.3\textwidth]{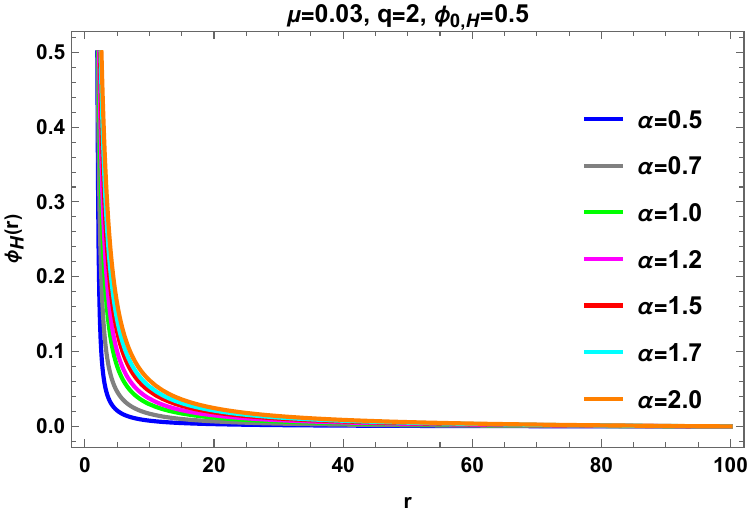}
\hfill%
\includegraphics[width=0.3\textwidth]{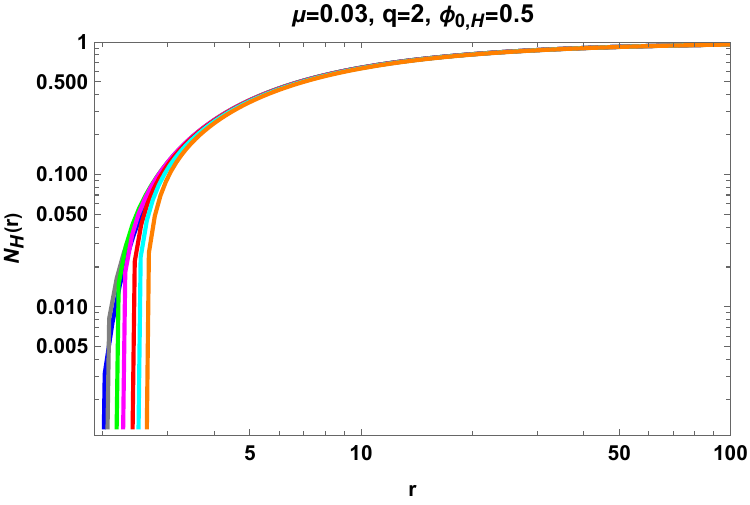}
\hfill%
\includegraphics[width=0.3\textwidth]{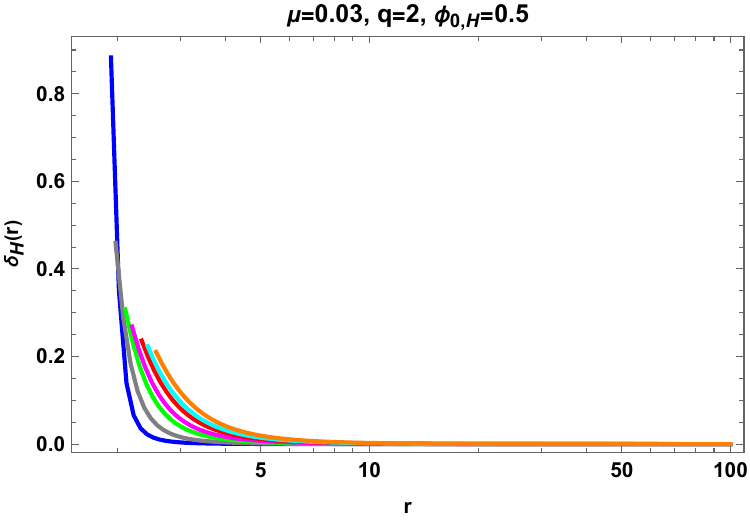}
\caption{Radial profile of scalar field $\phi_i(r)$, metric functions $N_i(r)$ and $\delta_i(r)$ with $\phi_{0,i}=0.5$ and seven  coupling constants $\alpha^i$ for each  $i=L,C,H$. From top to bottom, they  correspond to the scalarized  L-horizon ($q_L=0.5,q_{s,L}=0.6, 0.65, 0.68, 0.7, 0.74, 0.87, 0.97)$, C-horizon ($q_C=1,q_{s,C}=0.19, 0.22, 0.28, 0.34, 0.38, 0.47, 0.54$), and H-horizon ($q_H=2,q_{s,H}=0.07, 0.15, 0.27, 0.34, 0.43, 0.49, 0.56$) solutions, respectively. }\label{fig7}
\end{figure*}

Imposing two shooting conditions of $\delta_i(r\to\infty)=0$ and $\phi_i(r\to\infty)=0$, we obtain the fundamental ($n=0$) branch of the scalarized i-horizons.
Fig.~\ref{fig7} presents the profiles of the scalar hairs $\phi_i(r)$ and metric functions $N_i(r)$ and $\delta_i(r)$ for scalarized L, C, and H-horizon solutions corresponding to different values of each $\alpha^i$, with $q=0.5, 1$, and $2$, respectively.  Here, we fix the EH parameter $\mu=0.03$ and the horizon scalar $\phi_{0,i}=0.5$.
It can be seen that the scalar field $\phi_i(r)$ condenses to a nonzero value near the horizon and rapidly decays to zero as the radial coordinate $r$ increases. 
The metric function $N_i(r)$ vanishes at the horizon, then increases rapidly with increasing $r$, and asymptotically approaches unity at spatial infinity. 
Meanwhile, $\delta_i(r)$ also decays from a nonzero value to zero, indicating that the spacetime returns to an asymptotically flat metric.

It is worth emphasizing that the scalarzied i-horizon solutions corresponding to different $q$ exhibit distinct features and behaviors. 
As $q$ increases, the scalar field $\phi_i(r)$ decays more rapidly, which explains why the scalar profile becomes noticeably steeper near the horizon for $q=2$. 
In contrast, the metric function $N_i(r)$ grows more slowly with increasing $q$, such that for $q=0.5$,  $N_i(r)$ appears much steeper in the near-horizon.
For the L-horizon solution, an interesting feature is that $\delta_L(r)$ at the horizon remains nearly unchanged, while $\alpha$ mainly shifts the horizon location. 
On the other hand, for the C-horizon and H-horizon solutions, increasing $\alpha$ clearly reduces  $\delta_i(r)$ at the horizon with the decrease being significantly more pronounced for the H-horizon solution than for the C-horizon case.
Moreover, for all three types of solutions, we observe that increasing $\alpha^i$ leads to larger horizon radius $r_i$.

\section{Thermodynamic analysis}

In this section, we proceed to analyze the thermodynamic behavior of the hairy black hole solutions in different $q_i$ regions. 
The Hawking temperature and the area of i-horizon are defined as
\begin{equation}
    T_i=\frac{N_i'(r_i)}{4\pi}e^{-\delta_i(r_i)}, \quad a_{h,i}=\pi r^2_i.
\end{equation}

\begin{figure*}[t!]
\centering
\includegraphics[width=0.3\textwidth]{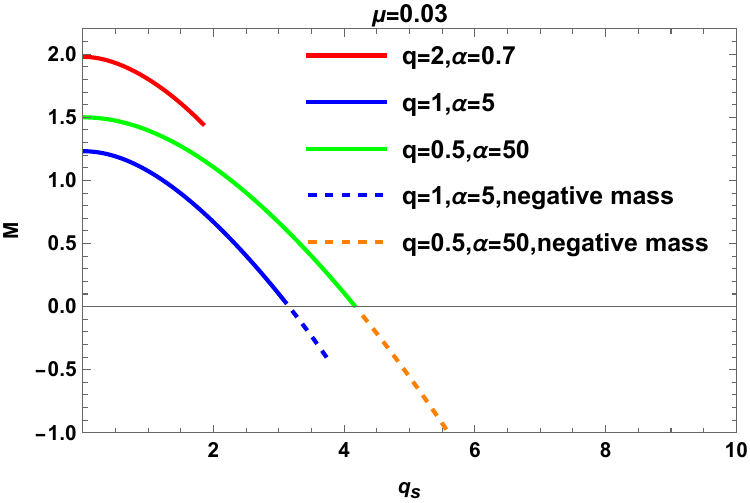}
\hfill%
\includegraphics[width=0.3\textwidth]{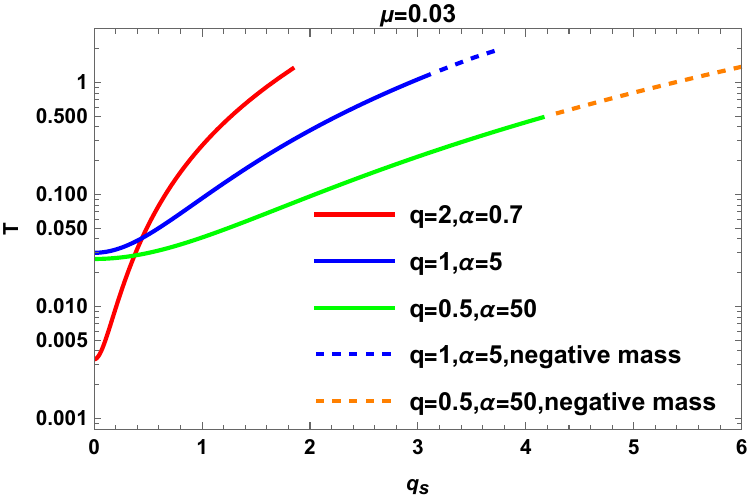}
\hfill%
\includegraphics[width=0.3\textwidth]{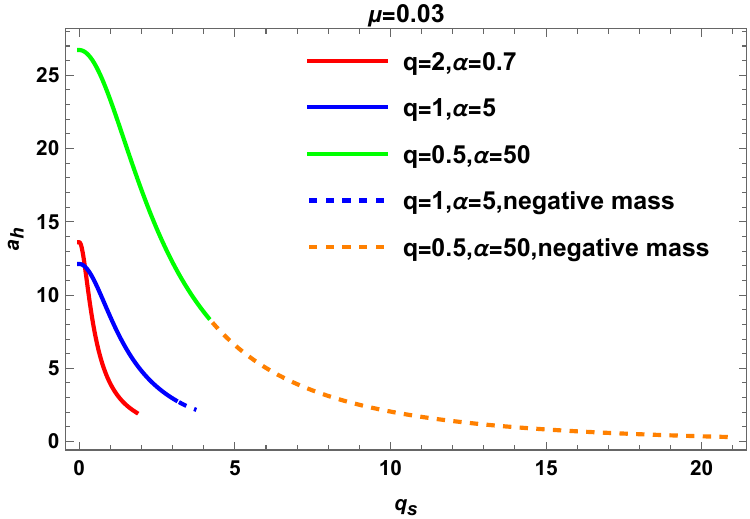}
\caption{(left) Black hole mass $M_i$, (middle) Hawking temperature $T_i$,  and (right) horizon area $a_{h,i}$ as  functions  of the scalar charge $q_{s,i}$ for L-horizon ($q_L=0.5,\alpha^L=50$), C-horizon ($q_C=1,\alpha^C=5$), and H-horizon ($q_H=2,\alpha^H=0.7$) solutions, respectively.}\label{fig8}
\end{figure*}

It is observed from Fig.~\ref{fig7} that since varying $\alpha^i$ causes quantitative differences but leaving the qualitative behavior unchanged, we may choose a representative set $\{\alpha^i\}$ to discuss the thermodynamic aspects of scalarized i-horizons: $q_L=0.5$ in [0,0.95] and $\alpha^L=50$ for the L-horizon solutions, $q_C=1$ in [0.95,1.0065] and $\alpha^C=5$ for the C-horizon solutions, and $q=2$ in [$0.95,\infty$] and $\alpha^H=0.7$ for the H-horizon solutions. 
We note that since $\alpha^i$ differs significantly across the different $q_i$ regions, a direct comparison between different $q_i$ cases is not meaningful.  
The above choices are made solely for our illustrative purpose.

\begin{figure*}[t!]
\centering
\includegraphics[width=0.4\textwidth]{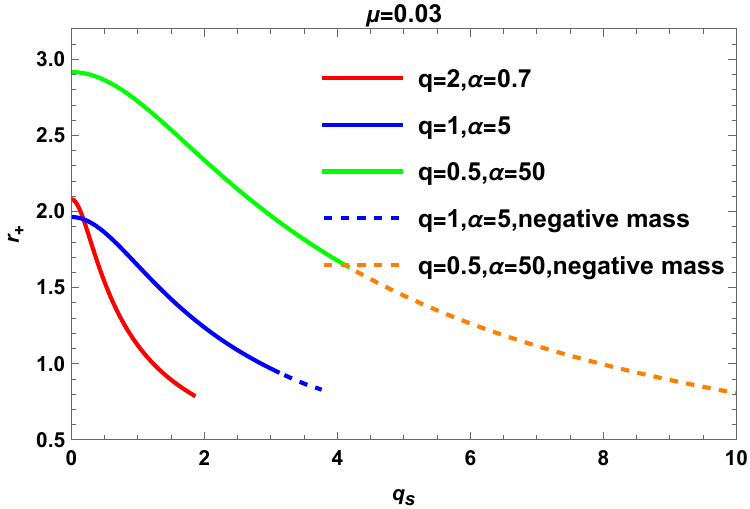}
\includegraphics[width=0.4\textwidth]{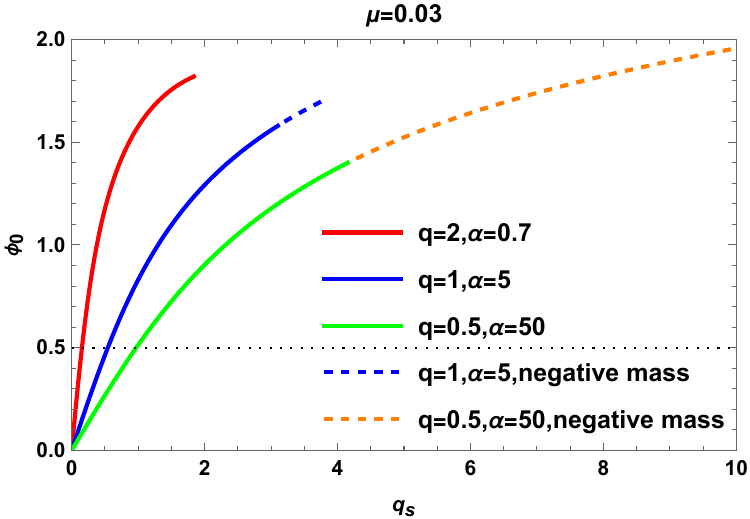}
\caption{Horizon radius $r_i$ (left) and horizon scalar  $\phi_{0,i}$ (right) as functions of the scalar charge $q_{s,i}$ for L-horizon ($q=0.5,\alpha=50$), C-horizon ($q=1,\alpha=5$), and H-horizon ($q=2,\alpha=0.7$) solutions, respectively. The black dotted horizontal line in the right figure represents $\phi_{0,i}=0.5$ at $q_s=q_{s,L}=0.97$, $q_{s,C}=0.54$, and $q_{s,H}=0.15$ as we discuss before.}\label{fig9}
\end{figure*}

The relevant results are presented in Fig.~\ref{fig8}. 
From the analysis of the black hole mass $M_i$ (left panel), we find that the black hole mass decreases monotonically as the scalar charge $q_s=q_{s,i}$ increases. 
An important feature is that, for both the C- and L-horizon solutions, a further increase of $q_s$ drives the black hole mass to negative values eventually. 
It indicates an emergence of unphysical solutions, where $M_i<0$ is marked by dashed lines in the figure. 
This determines the upper bounds on $q_{s,i}<q_{s,i}^u$ for $i=L,C$. 
This may arise  because for $M=1$, the allowed regions for L- and C-horizons are confined to be $q_L\in[0,0.95]$ and $q_C\in[0.95,1.0065]$, respectively.  
For a fixed $q=q_i$ in scalarized i-horizons, this confinement condition can be translated into the upper bounds on scalar charge $q_{s,i}$ through $M_i(q_{s,i})$. 

From the middle and right panels in Fig.~\ref{fig8}, one observes that the Hawking temperature $T_i$ increases monotonically with increasing scalar charge $q_{s,i}$, whereas the horizon area $a_{h,i}$ decreases monotonically. 
Combined with the behavior of the black hole mass $M_i$, this implies that high temperature is not allowed for $q_{s,i}>q^u_{s,i}$ with the scalarized L- and C-horizon solutions. 
Meanwhile, the horizon area (entropy) remains nonzero when the black hole mass approaches zero, although it tends to zero as $q_{s,i}$ increases further.  
This implies the lower entropy is not allowed for $q_{s,i}>q^u_{s,i}$ with the scalarized L and C-horizon solutions. 
For $q_{s,i}>q_{s,i}^u$ with $i=L,C$, thermodynamics variables (temperature, entropy) of scalarized i-horizons become unphysical. 

Similarly, Fig.~\ref{fig9} shows allowed regions for the horizon radius $r_i(q_s)$ and horizon scalar $\phi_{0,i}(q_s)$.  
We observe that the H-horizon radius decreases gradually with increasing scalar charge.  
Therefore, the growth of scalar charge $q_s$ leads to a monotonic decrease with respect to horizon radius.  
For the H-horizon scalar, increasing $q_s$ causes it to grow monotonically.  
However, for the scalarized L- and C-horizon solutions, a further increase of $q_s$ gives rise to excessively small horizon radius and large horizon scalar.  
They eventually lead to unallowable regions for negative black hole mass.

Consequently, the appearance of the upper bound ($q_{s,i}<q_{s,i}^u$) is considered as a new feature for scalarized L and C-horizon solutions. 

\section{Stability analysis  for the  scalarized i-horizons }

First of all, it is worthy mentioning that the stability analysis for scalarized i-horizons in the fundamental ($n=0$) branches is an important issue since it determines their viability to represent realistic astrophysical configurations.
Here, we wish to introduce the radial perturbations around the scalarized i-horizons as
\begin{eqnarray}
&&ds_{rp,i}^2=-N_i(r)e^{-2\delta_i(r)}(1+\epsilon H_{0,i})dt^2+\frac{dr^2}{N_i(r)(1+\epsilon H_{1,i})}
+r^2(d\theta^2+\sin^2\theta d\hat{\varphi}^2),\nonumber\\
&&\phi_i(t,r)=\phi_i(r)+\epsilon\delta\tilde{\phi_i}(t,r), \label{p-metric}
\end{eqnarray}
where $N_i(r)$, $\delta_i(r)$, and $\phi_i(r)$ represent the scalarized i-horizon background, whereas $H_{0,i}(t,r)$, $H_{1,i}(t,r)$, and $\delta\tilde{\phi_i}(t,r)$ denote three perturbations around the scalarized i-horizons.
Here, we do not need to consider a perturbation for the gauge field $A_{\hat{\varphi}}$.
Hereafter, we consider only the $l=0$(s-mode) scalar propagation by neglecting $l\neq0$. 
In this case, two metric perturbations become redundant ones.

Making a decoupling process, we are able to find a linearized scalar equation around the scalarized i-horizons.
Choosing the separation of variables
\begin{eqnarray}
\delta\tilde{\phi_i}(t,r)=\frac{\tilde{\varphi_i}(r)e^{\Omega t}}{r},
\end{eqnarray}
we obtain the Schr\"odinger-type equation
\begin{eqnarray}
\frac{d^2\tilde{\varphi_i}(r)}{dr_*^2}-\Big[\Omega_i^2+V_{s,i}(r,q_i,\phi_{0,i},\alpha^i)\Big]\tilde{\varphi_i}(r)=0
\end{eqnarray}
with $r_*$ is the tortoise coordinate defined by
\begin{eqnarray}
\frac{dr_*}{dr}=\frac{e^{\delta_i(r)}}{N_i(r)}.
\end{eqnarray}
Here, its potential is given by
\begin{eqnarray} 
&&V_{s,i}(r,M_i,q_i,\phi_{0,i},\alpha^i) \nonumber \\
&&\quad\quad\quad  =N_ie^{-2\delta_i}\Bigg[\frac{1-N_i-2r^2\phi_i'^2}{r^2}
-\frac{q_i^2\{1+\alpha^i(1+\phi_i^2+2r\phi_i^2\phi_i'^2)-2(-\alpha^i\phi_i+r\phi_i')^2\}}{r^4} \nonumber \\  
&&\quad\quad\quad + \frac{2\mu q_i^4\{1-r^2\phi'^2\}}{r^8} \Bigg]. \label{sc-poten}
\end{eqnarray}
We check easily that $V_{s,i}(r,M_i,q_i,\phi_{0,i},\alpha^i)$ with $\delta_i(r)=\phi_i(r)=0$ and $N_i(r)=f(r)$ reduces to the scalar potential around the EEHBH $V_{\rm EES}(r,M,q,\alpha)$ in Eq.(\ref{EEH-P}). 
It is worthy to note that horizon scalar $\phi_{0,i}$ is included as a variable instead of the scalar charge $q_{s,i}$ because the former is dependent on the latter as shown in (right) Fig. 9.

\begin{figure*}[t!]
\centering
\includegraphics[width=0.35\textwidth]{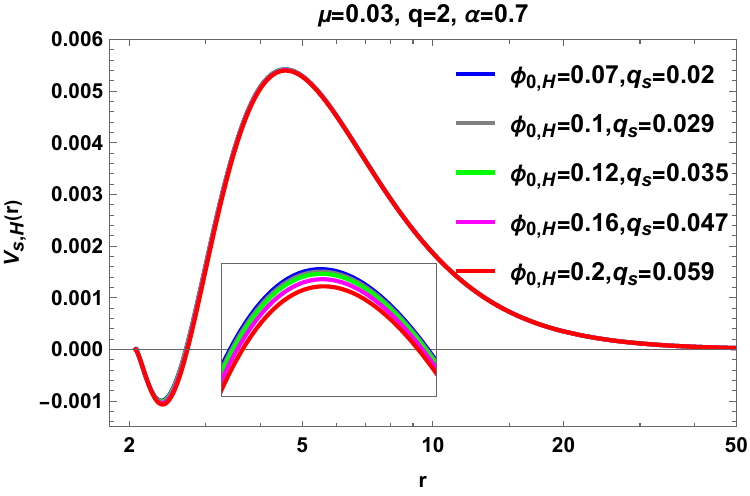}
\includegraphics[width=0.35\textwidth]{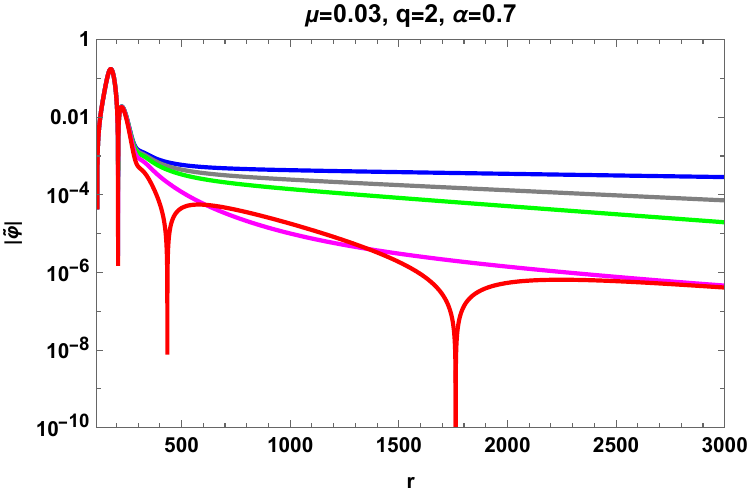}
\includegraphics[width=0.35\textwidth]{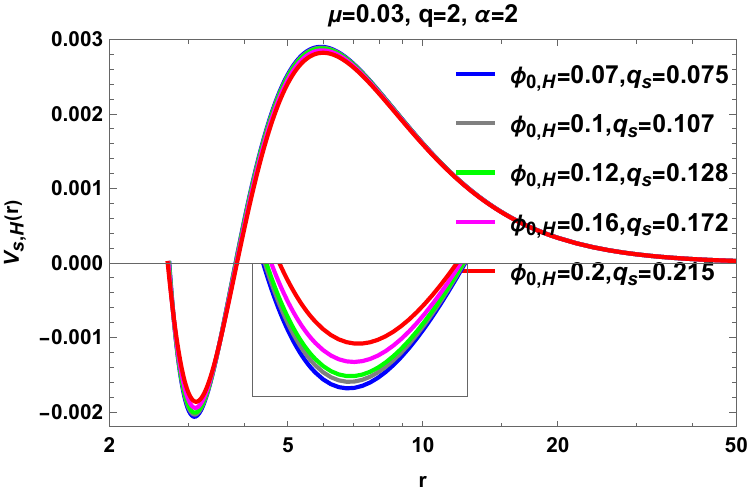}
\includegraphics[width=0.35\textwidth]{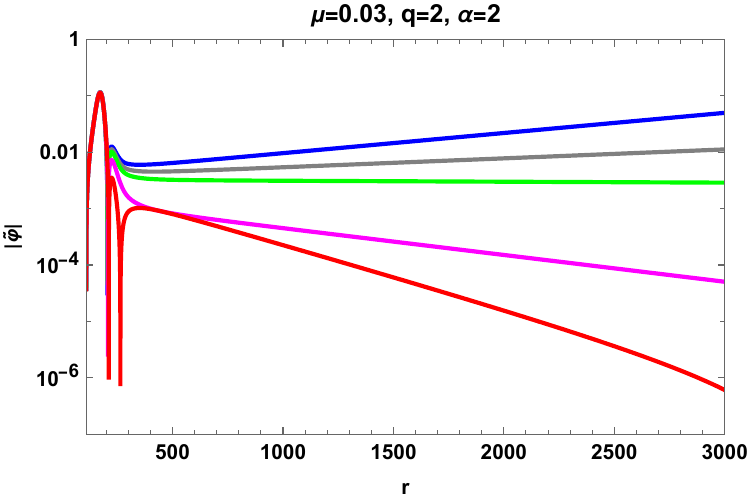}
\caption{Profile of the effective potential $V_{s,H}$ and corresponding time evolution of the scalar perturbation for scalarized H-horizon solutions for $\alpha=0.7$ (top) and $\alpha=2$ (bottom) with seven small horizon scalars $\phi_{0,H}$. }\label{fig10}
\end{figure*}

\begin{figure*}[t!]
\centering
\includegraphics[width=0.35\textwidth]{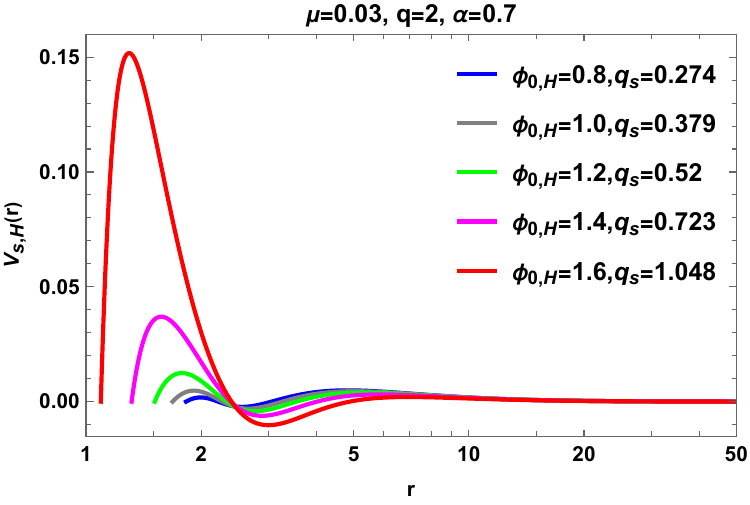}
\includegraphics[width=0.35\textwidth]{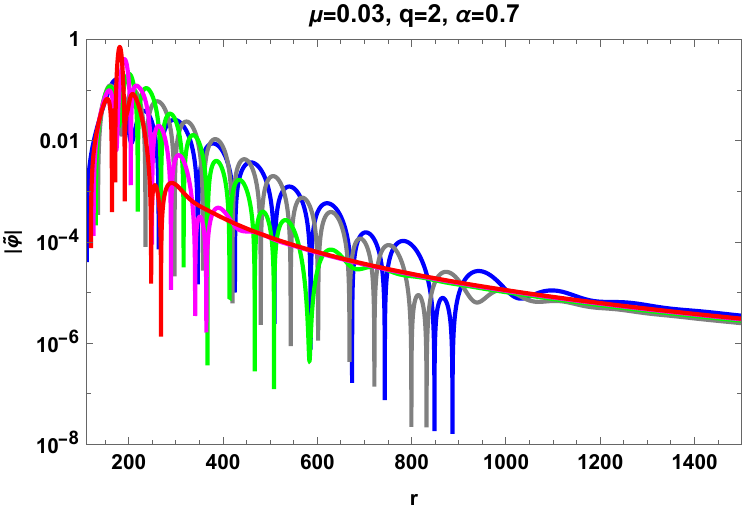}
\includegraphics[width=0.35\textwidth]{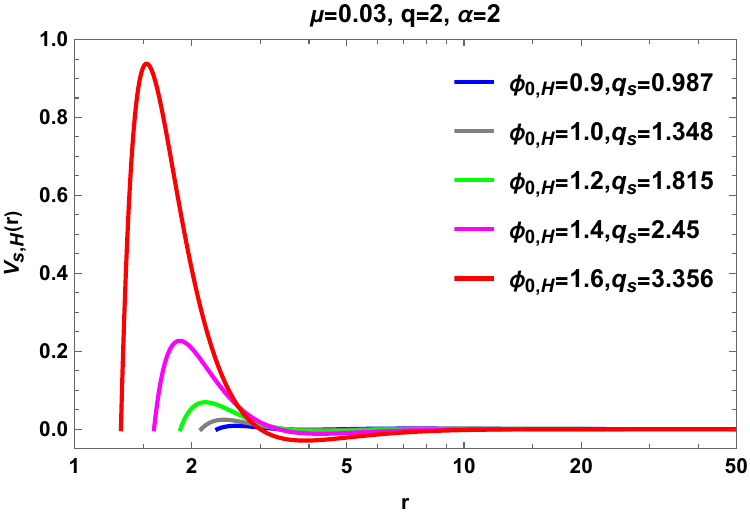}
\includegraphics[width=0.35\textwidth]{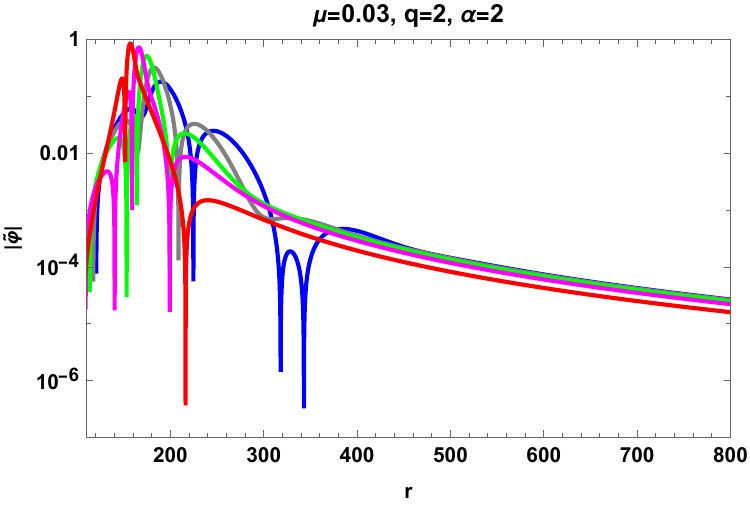}
\caption{Profile of the effective potential $V_{s,H}$ and corresponding time evolution of the scalar perturbation for scalarized  H-horizon solutions for $\alpha=0.7$ (top) and $\alpha=2$ (bottom) with seven large horizon scalars $\phi_{0,H}$.}\label{fig11}
\end{figure*}

\begin{figure*}[t!]
\centering
\includegraphics[width=0.35\textwidth]{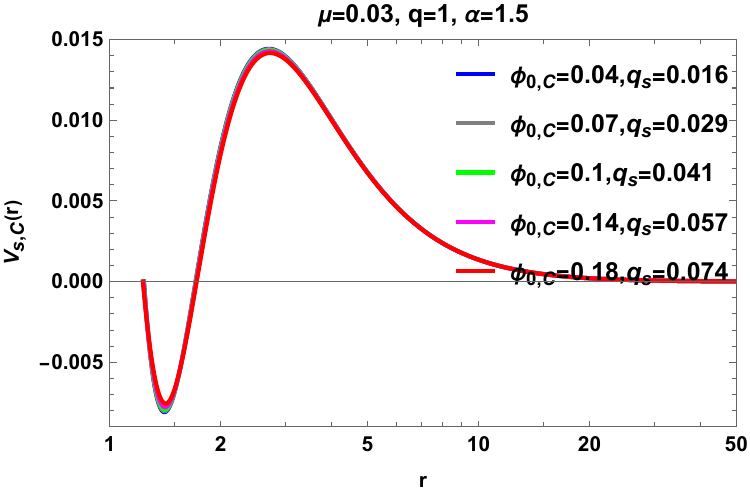}
\includegraphics[width=0.35\textwidth]{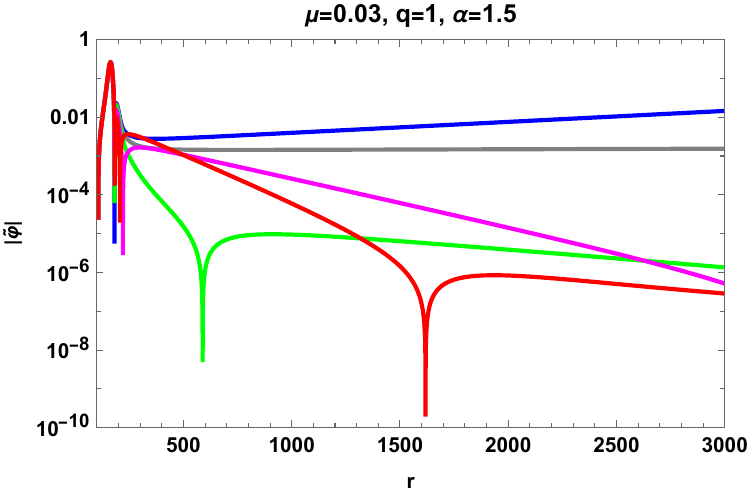}
\includegraphics[width=0.35\textwidth]{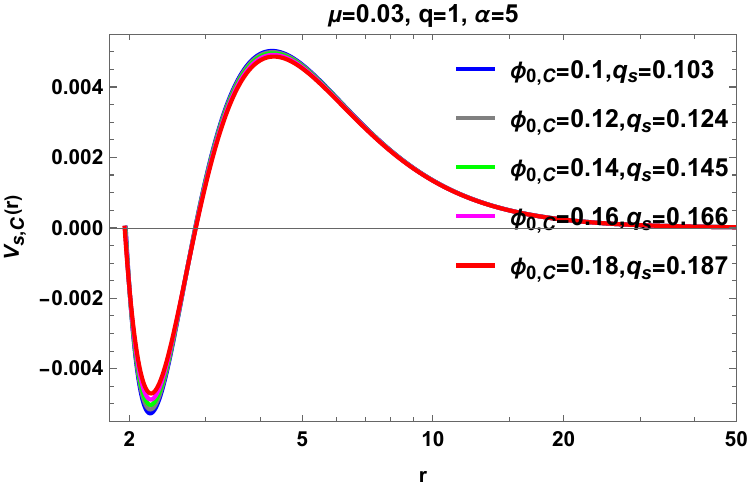}
\includegraphics[width=0.35\textwidth]{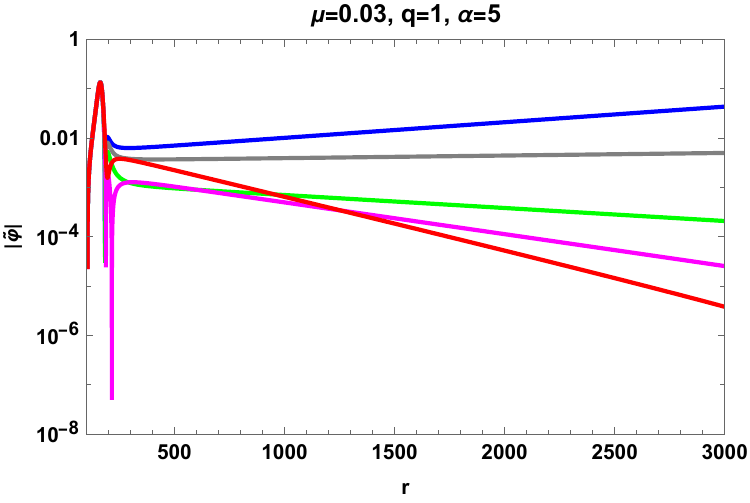}
\caption{Profile of the effective potential $V_{s,C}$ and corresponding time evolution of the scalar perturbation for scalarized C-horizon solutions for $\alpha=1.5$ (top) and $\alpha=5$ (bottom) with seven small horizon scalars $\phi_{0,C}$. }\label{fig12}
\end{figure*}

\begin{figure*}[t!]
\centering
\includegraphics[width=0.35\textwidth]{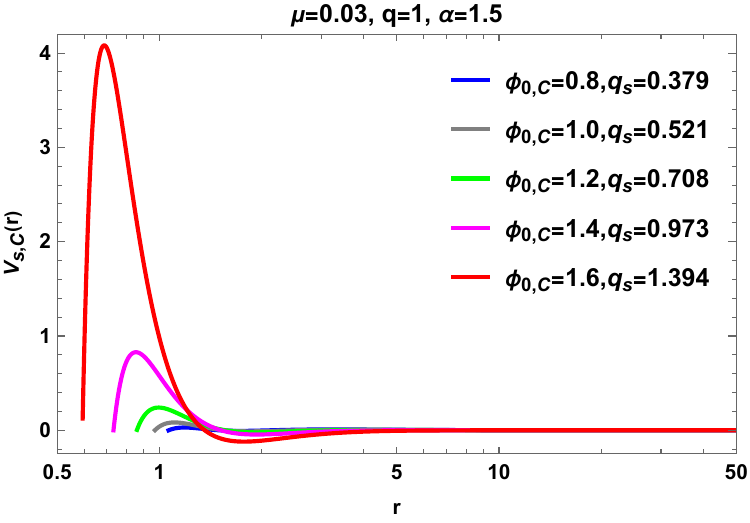}
\includegraphics[width=0.35\textwidth]{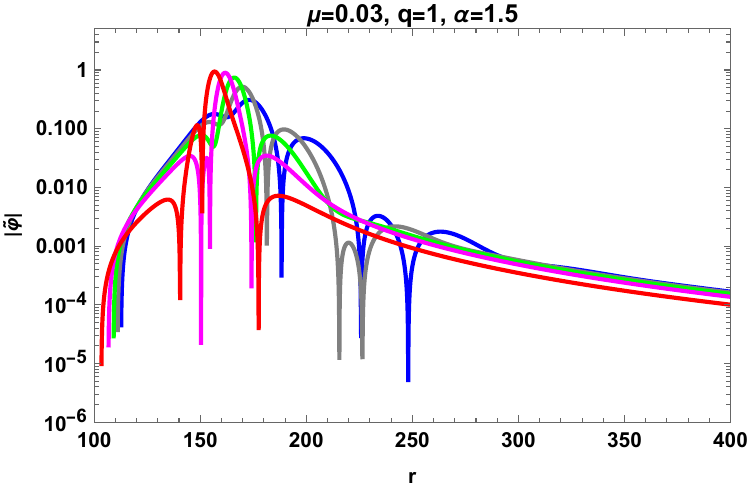}
\includegraphics[width=0.35\textwidth]{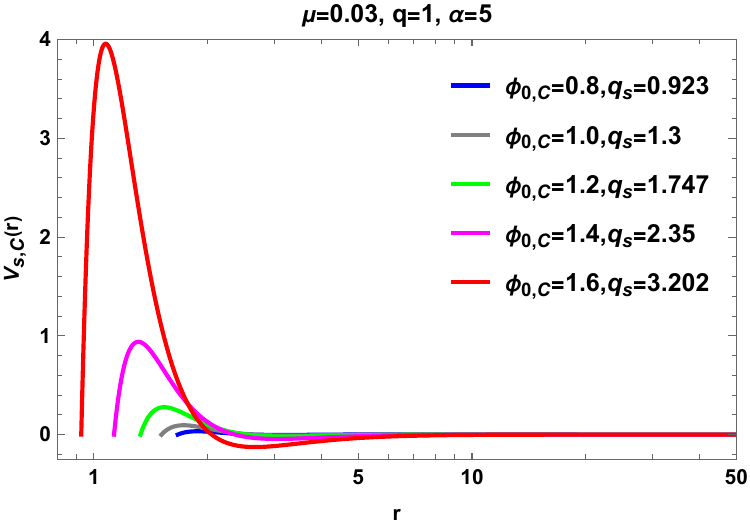}
\includegraphics[width=0.35\textwidth]{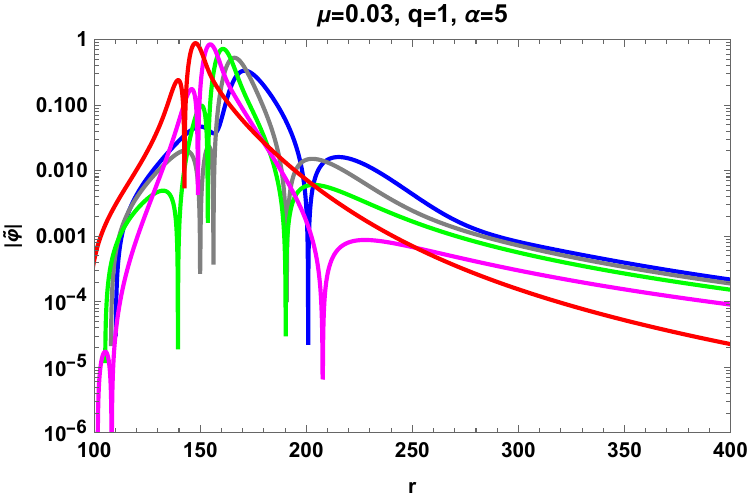}
\caption{Profile of the effective potential $V_{s,C}$ and corresponding time evolution of the scalar perturbation for scalarized  C-horizon solution for $\alpha=1.5$ (top) and $\alpha=5$ (bottom) with seven large horizon scalars $\phi_{0,C}$.}\label{fig13}
\end{figure*}

Since we encounter some difficulty in showing the stability issues of scalarized H-, C-, and  L-horizon solutions by computing QNM frequency of the scalar perturbations, we analyze their stability by computing the time evolution of scalar perturbations. 
Therefore, we discretize the perturbation equation as
\begin{align}\label{discre}   \frac{\tilde{\varphi}_i^{m,n+1}-2\tilde{\varphi}_i^{m,n}+\tilde{\varphi}_i^{m,n-1}}{\Delta r_*^2}-V_{s,i}^n\tilde{\varphi}_i^{m,n}=\frac{\tilde{\varphi}_i^{m+1,n}-2\tilde{\varphi}_i^{m,n}+\tilde{\varphi}_i^{m-1,n}}{\Delta t^2}+\mathcal{O}(\Delta t^2)+\mathcal{O}(\Delta r_*^2),
\end{align}
where $\tilde{\varphi}_i(t, r_*)=\tilde{\varphi}_i(m\Delta t,n\Delta r_*)=\tilde{\varphi}_i^{m,n}$ and $V_{s,i}=V_{s,i}(n\Delta r_*)=V_{s,i}^n$.
We consider an initial Gaussian wave packet $\tilde{\varphi}_i(t=0,r_*)=\exp(-\frac{(r_*-a)^2}{2b^2})$ with $a=10$ and $b=3$, and examine the time evolution of the scalar perturbation at $r=100$.
We perform numerical tests of the scalarized H-, C-, and L-horizons under different parameter choices. 
The corresponding effective potential profiles together with the associated waveform evolutions are presented in Figs.~\ref{fig10}–\ref{fig15}.

\subsection{H-horizon stability analysis}

For scalarized H-horizon with $q=2$, we consider two representative values ($\alpha = 0.7,~\alpha = 2$). 
In the small-$q_{s,H}$ regime shown in Fig.~\ref{fig10}, we observe some interesting features. 
For a relatively small coupling $\alpha=0.7$ [top of Fig.~\ref{fig10}], increasing the scalar charge $q_{s,H}$ does not trigger any instability and thus,   this background remains stable and the scalar perturbations decay more rapidly. 
This behavior is consistent with the shape of the  potential: increasing $q_{s,H}$ lowers the height of the potential barrier; although a potential well exists near the horizon, it is not sufficiently deep to induce any growth or divergence of the perturbations. 
For $\alpha=2$ [bottom of Fig.~\ref{fig10}], however, the potential develops a sufficiently deep well and the scalar perturbations exhibit exponential growth at small $q_{s,H}$.
This indicates that the scalarized H-horizon solution at small $q_{s,H}$ is unstable.
As $q_{s,H}$ increases, the depth of the potential well becomes progressively shallower and the perturbations are gradually stabilized until no instability is observed.
Accordingly, the scalarized H-solution undergoes a transition from unstable to stable configurations and a critical value $q^{\rm crit}_{s,H}$ can be inferred from the constant late-time behavior of the scalar perturbation. 

In the large-$q_{s,H}$ regime shown in Fig.~\ref{fig11}  where the scalar hair is sufficiently strong, both $\alpha=0.7$ and $\alpha=2$ lead to a sufficiently high potential barrier near the horizon while the potential well becomes comparatively shallow. 
In this regime, no sign of instability of the scalar perturbations is observed and increasing $q_{s,H}$ further accelerates the decay of the perturbations. 
This demonstrates that the scalarized H-horizon solution remains stable under scalar perturbations in the large-$q_{s,H}$ regime.

\subsection{Stability analysis for C- and L-horizons}

For the scalarized C-horizon black hole with $q=1$, we choose two couplings of $\alpha=1.5$ and $\alpha=5$.
In the small-$q_{s,C}$ regime shown in Fig.~\ref{fig12}, unlike the scalarized H-horizon case, the potential develops a sufficiently deep well in the near-horizon regardless of the value of $\alpha$.
Accordingly, we observe that the scalar perturbations diverge for small $q_{s,C}$ and become increasingly stable as $q_{s,C}$ increases. 
After reaching the critical value $q^{\rm crit}_{s,C}$, the perturbations are changed into a decaying mode. 
It is also found that as $\alpha$ increases, the critical value of the horizon scalar $\phi_{0,C}$ separating the unstable and stable regimes is shifted  to larger values.
In the large-$q_{s,C}$ regime, as illustrated in Fig.~\ref{fig13}, the scalar perturbations exhibit exponential decay for both couplings, indicating that the corresponding scalarized C-horizon solutions remain stable under perturbations.

\begin{figure*}[t!]
\centering
\includegraphics[width=0.35\textwidth]{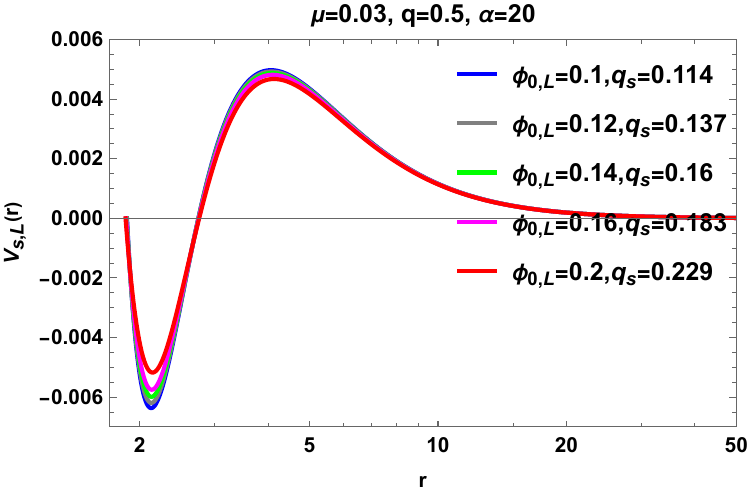}
\includegraphics[width=0.35\textwidth]{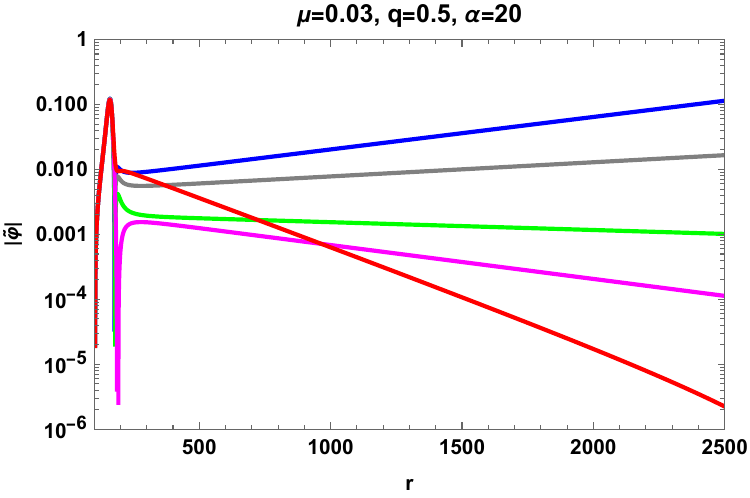}
\includegraphics[width=0.35\textwidth]{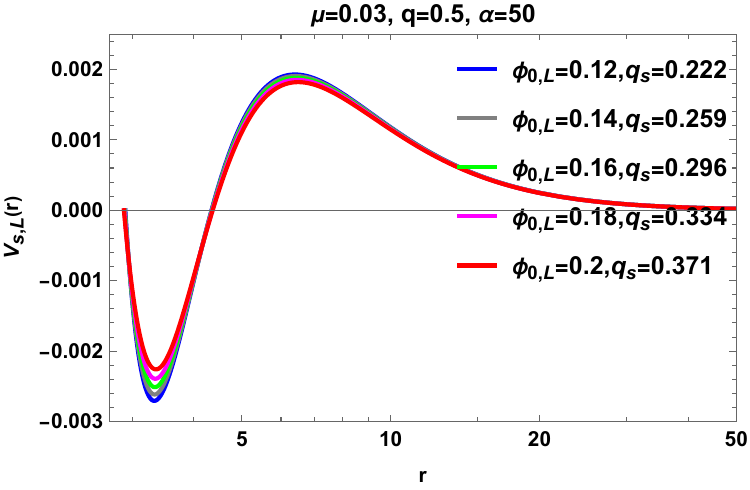}
\includegraphics[width=0.35\textwidth]{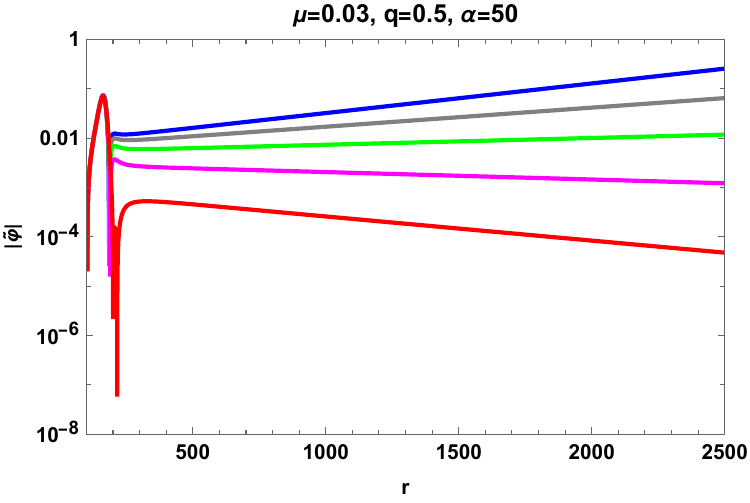}
\caption{Profile of the effective potential $V_{s,L}$ and corresponding time evolution of the scalar perturbation for scalarized L-horizon solution for $\alpha=20$ (top) and $\alpha=50$ (bottom) with seven small horizon scalars $\phi_{0,L}$. }\label{fig14}
\end{figure*}

\begin{figure*}[t!]
\centering
\includegraphics[width=0.35\textwidth]{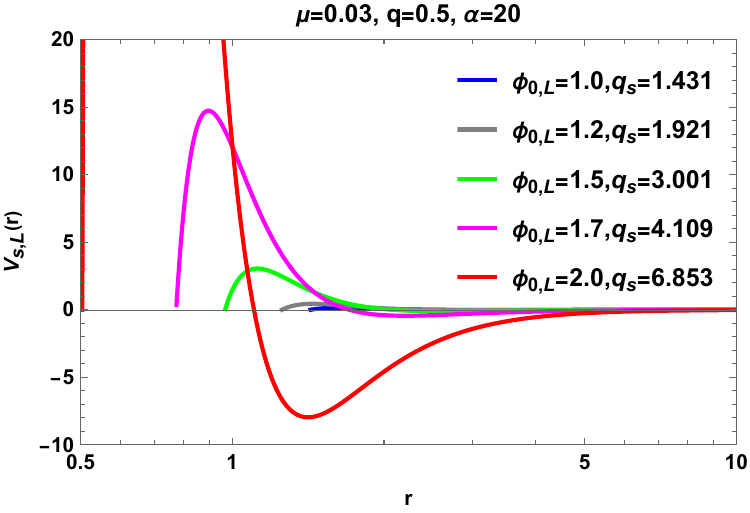}
\includegraphics[width=0.35\textwidth]{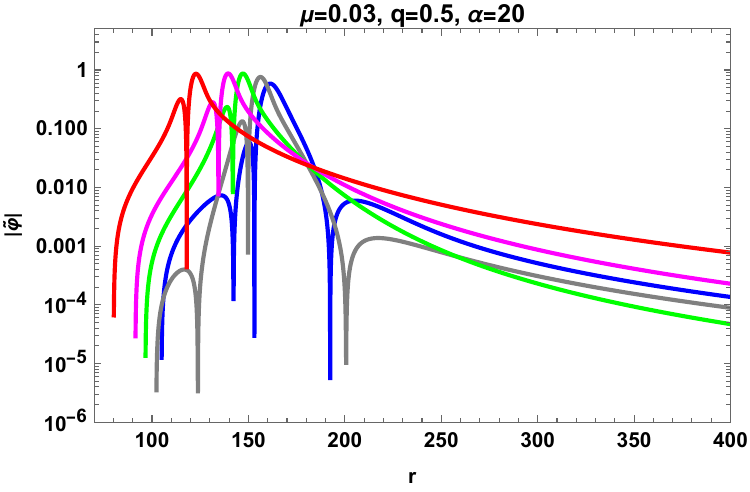}
\includegraphics[width=0.35\textwidth]{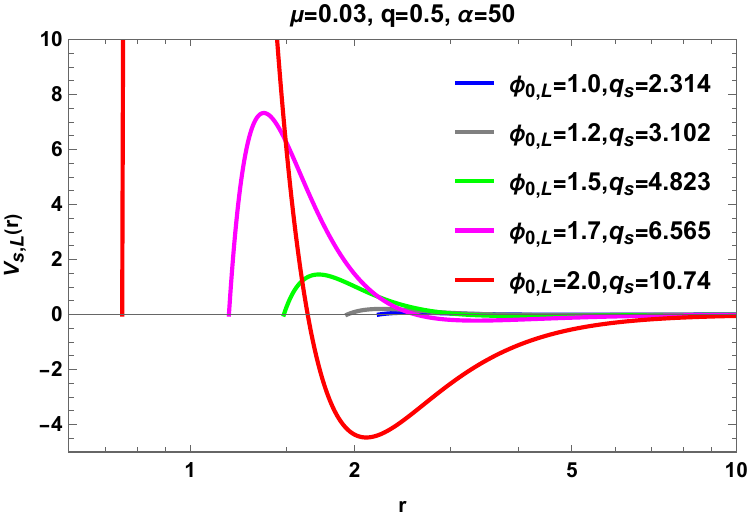}
\includegraphics[width=0.35\textwidth]{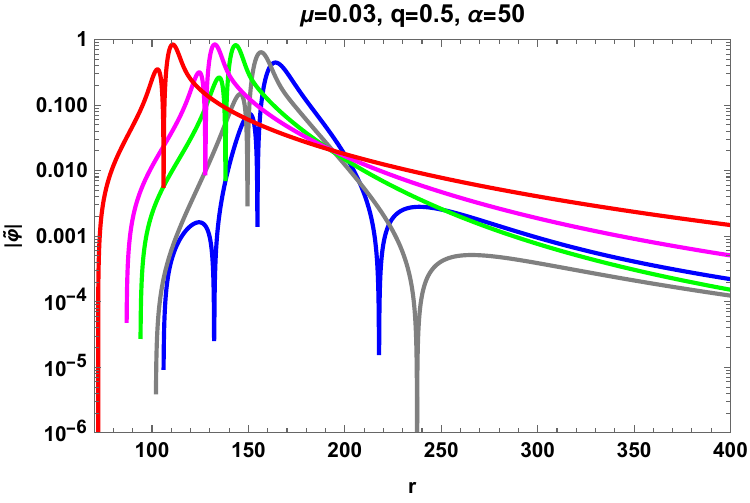}
\caption{Profile of the effective potential $V_{s,L}$ and corresponding time evolution of the scalar perturbation for scalarized L-horizon solution for $\alpha=20$ (top) and $\alpha=50$ (bottom) with seven large horizon scalars $\phi_{0,L}$.}\label{fig15}
\end{figure*}

For the scalarized L-horizon with $q=0.5$, we choose two couplings, $\alpha=20$ and $\alpha=50$. 
The scalarized L-horizon black holes exhibit qualitatively the same stability behavior as the scalarized C-horizon.
In the small-$q_{s,L}$ regime shown in Fig.~\ref{fig14}, the potential develops a sufficiently deep well in the near-horizon for both $\alpha=20$ and $\alpha=50$ and its depth decreases gradually as $q_{s,L}$ increases. 
As a result, the scalar perturbations diverge for small $q_{s,L}$ and transfer into a decaying behavior after reaching the critical value $q^{\rm crit}_{s,L}$. 
As in the C-horizon case, increasing $\alpha$ shifts the critical value of $\phi_{0,L}$ separating the unstable and stable regimes to larger values.
In the large-$q_{s,L}$ regime shown in Fig.~\ref{fig15}, the scalar perturbations exhibit exponential decay, implying that the scalarized L-horizon solution is stable under perturbations.

In summary of the scalarized H-, C-, and L-horizons, we find that all solutions remain stable under scalar perturbations in the large-$q_{s,i}$ regime, whereas instabilities arise only in the small-$q_{s,i}$ regime. 
For the H-horizon case, when $\alpha$ is relatively small, the black hole solutions remain stable at small $q_{s,i}$, and increasing $\alpha$ triggers unstable modes in the small-$q_{s,i}$ regime. 
For the scalarized C- and L-horizons, in contrast, unstable solutions in the small-$q_{s,i}$ regime are found for both the smaller and larger couplings considered here, which distinguishes them from the scalarized H-horizon where a small coupling yields no instability.
A unified picture underlying these behaviors is that the onset of instability is governed by the depth of the potential well in $V_{s,i}$ in the near-horizon. 

Finally, it is worth noting that the recent results on the negative potential-induced scalarization of the single-horizon EEHBH ($\mu=0.3$)~\cite{Guo:2025ksj} indicate that the single branch of hairy black holes is unstable against radial perturbations. 
However, the present work establishes that the instability is constrained by sufficiently large scalar charge in the multi-horizon regime ($\mu=0.03$) through spontaneous scalarization.

\subsection{New observations from time-evolution analysis}

The unstable-to-stable transition in the fundamental branches is regarded as a characteristic feature of the EEHS theory because such a pattern has never been observed in the known analyses. 
We note that for the EMS theory with quartic coupling, the cold branch with smaller scalar charge of nonlinear scalarized black holes is unstable, while the hot branch with larger scalar charge is stable against radial perturbations~\cite{Blazquez-Salcedo:2020nhs,LuisBlazquez-Salcedo:2020rqp}.
Moreover, scalarized black holes in EMS theory within AdS spacetime exhibit a similar phenomenon near the bifurcation line~\cite{Guo:2021zed}.
Also, the quadratic scalar coupling to the Gauss-Bonnet term yields unstable scalarized solutions, while stable solutions can be obtained after including higher-order self-interaction terms~\cite{Blazquez-Salcedo:2018jnn,Silva:2018qhn,Macedo:2019sem}. 
In contrast, the quadratic coupling of the Maxwell term in EMS theory leads to a stable fundamental branch of scalarized charged solutions through spontaneous scalarization~\cite{Myung:2019oua}.
To our knowledge, this transition in the fundamental branch found here has not been previously reported for asymptotically flat scalarized black holes.
Its appearance should be attributed to the NED term and its EEHBH solution. 

Another observation to be addressed is that in the large-$q_{s,i}$ regime, the black hole mass for the  scalarized C- and L-horizon cases becomes negative, which renders these solutions thermodynamically unstable.  
Nevertheless, the dynamical stability analysis shows that these black hole solutions remain stable under scalar perturbations.
We stress that these two analyses are completely independent and probe different notions of viability.
In this sense, the two analyses are considered as complementary.
The dynamical instability excludes the small-$q_{s,i}$ end of each branch ($q_{s,i}<q^{\rm crit}_{s,i}$), while the positivity condition for the mass removes its large-$q_{s,i}$ tail ($q_{s,i}>q^u_{s,i}$, present only for $i=L,C$). 
Consequently, physically acceptable and dynamically stable scalarized black holes exist only in an intermediate window for scalarized L- and C-horizon solutions.
The appearance of this two-sided window on the primary scalar charge indicates a novel feature of scalarized EEHBHs with multiple horizons.

A further observation states that as $q_{s,i}$ increases, the height of the potential barrier undergoes dramatic changes, spanning several orders of magnitude from approximately $10^{-3}$ to $\mathcal{O}(10^{2})$. 
In the large-$q_{s,i}$ regime, the resulting extremely high and sharp potential barrier leads to a very rapid decay of the perturbations, such that only a few oscillations are visible in Figs.~\ref{fig11},~\ref{fig13}, and~\ref{fig15} before the signal effectively disappears. 
In the small-$q_{s,i}$ regime, exponentially growing perturbations exhibit no oscillatory behavior as well as exponentially decaying modes are also non-oscillatory. 
This implies that the corresponding quasinormal mode (QNM) frequencies are purely imaginary. 
As $q_{s,i}$ increases further, one observes that oscillatory behavior gradually emerges, leading to complex QNM frequencies.
Therefore, the time-domain evolution suggests strongly that the computation of QNM frequencies in this regime is considered as a numerically challenging task compared to our previous work in Ref.~\cite{Guo:2025ksj}. 

\section{Discussions }

In this work, we have investigated scalarized black holes with multiple horizons in the EEHS theory by introducing a quadratic scalar coupling $g_i(\phi)=1-\alpha^i\phi^2$ to the Maxwell term.
This extends previous studies of scalarized EEHBHs with single horizon ($\mu=0.3$)~\cite{Guo:2025ksj,Zhang:2025msi,Zhang:2026bqu} to the multi-horizon regime ($\mu=0.03\le 0.08$),  whose horizon structure plays a nontrivial role in the scalarization analysis.

For $M=1$ and $\mu=0.03$, the condition of $f(r)=0$ yields four distinct horizon families $r_L$, $r_C$, $r_N$, and $r_H$.
The subscripts L, C, H, and N carry clear thermodynamic meanings: the cold (C) horizon branch is thermodynamically stable throughout its band, the low (L) horizon branch exhibits a Davies point at $q=0.871$ where a rapid phase transition occurs, the hot (H) horizon branch contains a thermodynamically unstable region, and the negative (N) horizon contains negative temperature and is closely analogous to the inner horizon of the RNBH. 
A triple horizon configuration ($r_H<r_N<r_C$) exists only in a narrow band $q\in[0.95,1.0065]$ bounded by the merging point of L- and C-horizons at $q=0.95$ and the extremal point at $q=1.0065$. 
In contrast to the single horizon case with $\mu>0.08$, the L-horizon is disconnected from the H-horizon. 
Notably, the L-horizon reproduces the RN outer horizon in both its geometry and its instability boundary [$\alpha_{s,L}(1,q_L,0.03)\approx \alpha_{sRN}(1,q)$], indicating that the NED correction is negligible for small charge. 
However, the L-horizon does not contain any extremal point.

We are interested in  the onset analysis of scalarization carried out around each horizon. 
The sufficient condition $I_i<0$ for tachyonic instability determines $\alpha_{s,i}(M,q_i,\mu)$. 
The WKB approximation yields the instability bounds $\alpha^i_{{\rm in},n}(M,q_i,\mu)$ for infinite  branches of $n=0,1,2,\cdots$.
Each of the low, cold, and high horizons admits infinite branches of scalarized EEHBH solutions, closely resembling the RNBH in the EMS theory. 
However, a distinctive feature of the triple horizons emerges: the WKB integrations for the hot and negative horizons are not properly defined because their metric function $f_i(r)$ becomes negative in the near-horizon outside $r_H$ and $r_N$.
As a consequence, within the narrow band $q\in[0.95,1.0065]$, the cold horizon represents the most reliable reference background for scalarization analysis, while scalar clouds associated with the negative horizon are not well-defined, suggesting the absence of  a meaningful scalarization.

Based on the onset analysis, we have constructed the three fundamental ($n=0$) branches of scalarized L-, C-, and H-horizons by solving the full equations with two shooting conditions [$\delta_i(\infty)=0$ and $\phi_i(\infty)=0$]. 
The scalarzied i-horizon solutions corresponding to different $q=0.5,1,2$ exhibit distinct features and behaviors. 
The scalar hair $\phi_i(r)$ decays more rapidly as $q$ increases, which explains why the scalar profile becomes noticeably steeper in the near-horizon for the case of $q=2$. 
On the other hand, the metric function $N_i(r)$ grows more slowly with increasing $q$, such that for $q=0.5$ case, the profile of $N_i(r)$ appears much steeper in the near-horizon than others.
For all three types of solutions, we observed that increasing coupling constant $\alpha^i$ leads to larger horizon radius $r_i$.

Concerning thermodynamic analysis of the scalarized $i$-horizons, we found the existence of the upper bound on primary scalar charge $q_{s,i}$.
We observed that black hole mass $M_i$ decreases monotonically as the scalar charge $q_{s,i}$ increases. 

An important observed feature is that for both scalarized C- and L-horizon solutions, a further increase of $q_{s,i}$ drives the black hole mass $M_i$ to negative values eventually.
This indicates the upper bound of $q_{s,i}<q^u_{s,i}$.
It might arise because for $M=1$, the allowed regions for L- and C-horizon are confined to be $q_L\in[0,0.95]$ and $q_C\in[0.95,1.0065]$, respectively.
For fixed $q=0.5,1$ in scalarized i-horizons, the confinement condition can be translated into the upper bounds on the scalar charge as $q_{s,i}<q^u_{s,i}$ through  mass $M_i(q_{s,i})$. 
In this case, we found that high Hawking temperature (low entropy) renders the scalarized C-horizon and L-horizon solutions unphysical because their mass becomes negative. 

Finally, the time-domain analysis of radial perturbations showed that all three fundamental branches are stable in the large-$q_{s,i}$ regime, where a high potential barrier significantly enhances damping of perturbations.
However, instabilities arise only at small $q_{s,i}$. 
For the scalarized H-horizon solution, its instability is triggered by increasing $\alpha$, whereas for the scalarized L- and C-horizon solutions, sufficiently deep potential wells appear even at small $\alpha$. 
In the small-$q_{s,i}$ regime, scalar perturbations exhibit non-oscillatory growth or decay, corresponding to purely imaginary QNM frequencies. 
As $q_{s,i}$ increases, oscillatory behavior emerges, leading to complex QNM frequencies.
Therefore, in contrast with the negative potential-induced scalarization of the single-horizon EEHBH whose single branch is unstable~\cite{Guo:2025ksj}, the present spontaneous scalarization in the multi-horizon regime admits dynamically stable hairy black hole solutions within either a two-sided window on the primary scalar charge for scalarized L- and C-horizon solutions or a lower bound of primary scalar charge for the H-horizon solution.
 
 \vspace{1cm}

{\bf Acknowledgments}

 Y.S.M. was supported by the National Research Foundation of Korea (NRF) grant funded by the Korea government(MSIT) (RS-2022-NR069013).
 

\vspace{1cm}


\newpage
\begin{thebibliography}{99}
\bibitem{Ruffini:1971bza}
  R.~Ruffini and J.~A.~Wheeler,
  Phys.\ Today {\bf 24}, no. 1, 30 (1971).
  doi:10.1063/1.3022513

\bibitem{Bekenstein:1972ny}
J.~D.~Bekenstein,
Phys. Rev. Lett. \textbf{28}, 452-455 (1972)
doi:10.1103/PhysRevLett.28.452

\bibitem{Bekenstein:1971hc}
J.~D.~Bekenstein,
Phys. Rev. D \textbf{5}, 1239-1246 (1972)
doi:10.1103/PhysRevD.5.1239

\bibitem{Herdeiro:2015waa}
  C.~A.~R.~Herdeiro and E.~Radu,
  Int.\ J.\ Mod.\ Phys.\ D {\bf 24}, no. 09, 1542014 (2015)
  doi:10.1142/S0218271815420146
  [arXiv:1504.08209 [gr-qc]].


\bibitem{Bekenstein:1995un}
J.~D.~Bekenstein,
Phys. Rev. D \textbf{51}, no.12, R6608 (1995)
doi:10.1103/PhysRevD.51.R6608

\bibitem{Doneva:2017bvd}
D.~D.~Doneva and S.~S.~Yazadjiev,
Phys. Rev. Lett. \textbf{120}, no.13, 131103 (2018)
doi:10.1103/PhysRevLett.120.131103
[arXiv:1711.01187 [gr-qc]].

\bibitem{Silva:2017uqg}
H.~O.~Silva, J.~Sakstein, L.~Gualtieri, T.~P.~Sotiriou and E.~Berti,
Phys. Rev. Lett. \textbf{120}, no.13, 131104 (2018)
doi:10.1103/PhysRevLett.120.131104
[arXiv:1711.02080 [gr-qc]].

\bibitem{Antoniou:2017acq}
G.~Antoniou, A.~Bakopoulos and P.~Kanti,
Phys. Rev. Lett. \textbf{120}, no.13, 131102 (2018)
doi:10.1103/PhysRevLett.120.131102
[arXiv:1711.03390 [hep-th]].

\bibitem{Herdeiro:2018wub}
C.~A.~R.~Herdeiro, E.~Radu, N.~Sanchis-Gual and J.~A.~Font,
Phys. Rev. Lett. \textbf{121}, no.10, 101102 (2018)
doi:10.1103/PhysRevLett.121.101102
[arXiv:1806.05190 [gr-qc]].

\bibitem{Myung:2018vug}
Y.~S.~Myung and D.~C.~Zou,
Eur. Phys. J. C \textbf{79}, no.3, 273 (2019)
doi:10.1140/epjc/s10052-019-6792-6
[arXiv:1808.02609 [gr-qc]].

\bibitem{Heisenberg:1936nmg}
W.~Heisenberg and H.~Euler,
Z. Phys. \textbf{98}, no.11-12, 714-732 (1936)
doi:10.1007/BF01343663
[arXiv:physics/0605038 [physics]].



\bibitem{Obukhov:2002xa}
Y.~N.~Obukhov and G.~F.~Rubilar,
Phys. Rev. D \textbf{66}, 024042 (2002)
doi:10.1103/PhysRevD.66.024042
[arXiv:gr-qc/0204028 [gr-qc]].

\bibitem{Sorokin:2021tge}
D.~P.~Sorokin,
Fortsch. Phys. \textbf{70} (2022) no.7-8, 2200092
doi:10.1002/prop.202200092
[arXiv:2112.12118 [hep-th]].


\bibitem{Yajima:2000kw}
H.~Yajima and T.~Tamaki,
Phys. Rev. D \textbf{63}, 064007 (2001)
doi:10.1103/PhysRevD.63.064007
[arXiv:gr-qc/0005016 [gr-qc]].

\bibitem{Myung:2025zxu}
Y.~S.~Myung,
[arXiv:2503.18239 [gr-qc]].



\bibitem{Karakasis:2022xzm}
T.~Karakasis, G.~Koutsoumbas, A.~Machattou and E.~Papantonopoulos,
Phys. Rev. D \textbf{106}, no.10, 104006 (2022)
doi:10.1103/PhysRevD.106.104006
[arXiv:2207.13146 [gr-qc]].

\bibitem{Guo:2025ksj}
H.~Guo, M.~Park and Y.~S.~Myung,
Chin. Phys. C \textbf{50} (2026) no.6, 065102
doi:10.1088/1674-1137/ae457c
[arXiv:2508.16083 [gr-qc]].

\bibitem{Zhang:2025msi}
L.~Zhang, D.~C.~Zou and Y.~S.~Myung,
Eur. Phys. J. C \textbf{85}, no.12, 1463 (2025)
doi:10.1140/epjc/s10052-025-15232-4
[arXiv:2510.07954 [gr-qc]].

\bibitem{Zhang:2026bqu}
L.~Zhang, Y.~S.~Myung, D.~C.~Zou and C.~M.~Zhang,
[arXiv:2605.15566 [gr-qc]].

\bibitem{Amaro:2020xro}
D.~Amaro and A.~Mac\'\i{}as,
Phys. Rev. D \textbf{102}, no.10, 104054 (2020)
doi:10.1103/PhysRevD.102.104054

\bibitem{Breton:2021mju}
N.~Bret\'on and L.~A.~L\'opez,
Phys. Rev. D \textbf{104}, no.2, 024064 (2021)
doi:10.1103/PhysRevD.104.024064
[arXiv:2105.12283 [gr-qc]].



\bibitem{Allahyari:2019jqz}
A.~Allahyari, M.~Khodadi, S.~Vagnozzi and D.~F.~Mota,
JCAP \textbf{02}, 003 (2020)
doi:10.1088/1475-7516/2020/02/003
[arXiv:1912.08231 [gr-qc]].


\bibitem{Gao:2021kvr}
C.~Gao,
Phys. Rev. D \textbf{104}, no.6, 064038 (2021)
doi:10.1103/PhysRevD.104.064038
[arXiv:2106.13486 [gr-qc]].


\bibitem{Blazquez-Salcedo:2020nhs}
J.~L.~Bl{\'a}zquez-Salcedo, C.~A.~R.~Herdeiro, J.~Kunz, A.~M.~Pombo and E.~Radu,
Phys. Lett. B \textbf{806}, 135493 (2020)
doi:10.1016/j.physletb.2020.135493
[arXiv:2002.00963 [gr-qc]].

\bibitem{LuisBlazquez-Salcedo:2020rqp}
J.~Luis Bl{\'a}zquez-Salcedo, C.~A.~R.~Herdeiro, S.~Kahlen, J.~Kunz, A.~M.~Pombo and E.~Radu,
Eur. Phys. J. C \textbf{81}, no.2, 155 (2021)
doi:10.1140/epjc/s10052-021-08952-w
[arXiv:2008.11744 [gr-qc]].




\bibitem{Luo:2022gdz}
Z.~Luo and J.~Li,
Chin. Phys. C \textbf{46}, no.8, 085107 (2022)
doi:10.1088/1674-1137/ac6574




\bibitem{Dotti:2004sh}
G.~Dotti and R.~J.~Gleiser,
Class. Quant. Grav. \textbf{22}, L1 (2005)
doi:10.1088/0264-9381/22/1/L01
[arXiv:gr-qc/0409005 [gr-qc]].

\bibitem{Hod:2019ulh}
S.~Hod,
Phys. Lett. B \textbf{798} (2019), 135025
doi:10.1016/j.physletb.2019.135025
[arXiv:2002.01948 [gr-qc]].

\bibitem{Guo:2021zed}
G.~Guo, P.~Wang, H.~Wu and H.~Yang,
Eur. Phys. J. C \textbf{81} (2021) no.10, 864
doi:10.1140/epjc/s10052-021-09614-7
[arXiv:2102.04015 [gr-qc]].

\bibitem{Blazquez-Salcedo:2018jnn}
J.~L.~Bl{\'a}zquez-Salcedo, D.~D.~Doneva, J.~Kunz and S.~S.~Yazadjiev,
Phys. Rev. D \textbf{98} (2018) no.8, 084011
doi:10.1103/PhysRevD.98.084011
[arXiv:1805.05755 [gr-qc]].

\bibitem{Silva:2018qhn}
H.~O.~Silva, C.~F.~B.~Macedo, T.~P.~Sotiriou, L.~Gualtieri, J.~Sakstein and E.~Berti,
Phys. Rev. D \textbf{99} (2019) no.6, 064011
doi:10.1103/PhysRevD.99.064011
[arXiv:1812.05590 [gr-qc]].

\bibitem{Macedo:2019sem}
C.~F.~B.~Macedo, J.~Sakstein, E.~Berti, L.~Gualtieri, H.~O.~Silva and T.~P.~Sotiriou,
Phys. Rev. D \textbf{99} (2019) no.10, 104041
doi:10.1103/PhysRevD.99.104041
[arXiv:1903.06784 [gr-qc]].

\bibitem{Myung:2019oua}
Y.~S.~Myung and D.~C.~Zou,
Eur. Phys. J. C \textbf{79} (2019) no.8, 641
doi:10.1140/epjc/s10052-019-7176-7
[arXiv:1904.09864 [gr-qc]].

\end{thebibliography}
\end{document}